\documentclass[journal]{IEEEtran}
\usepackage{amssymb}
\usepackage{amsmath}
\usepackage{algorithm}
\usepackage{algorithmic}
\usepackage{multicol}
\usepackage{ragged2e}
\usepackage{subfigure}
\usepackage{lipsum}
\usepackage{mathtools,lipsum}

\usepackage{cuted}
\usepackage{graphicx,mwe}
\usepackage{multirow}
\usepackage{pifont}
\usepackage{rotating}
\usepackage{cite}
\usepackage{xcolor}
\usepackage{fancyhdr}
\begin{document}
	%
	\title{Energy Efficient UAV-Assisted Emergency Communication with Reliable Connectivity and Collision Avoidance}
	\author{N P~Sharvari,~\IEEEmembership{Student Member,~IEEE,}
		Dibakar~Das,~\IEEEmembership{Senior~Member,~IEEE,}
		Jyotsna~Bapat,~\IEEEmembership{Member,~IEEE,}
		and~Debabrata~Das,~\IEEEmembership{Senior~Member,~IEEE,}}
	\maketitle
	\thispagestyle{fancy}
	\begin{abstract}Emergency communication is vital for search and rescue operations following natural disasters. Unmanned Aerial Vehicles (UAVs) can significantly assist emergency communication by agile positioning, maintaining connectivity during rapid motion, and relaying critical disaster-related information to Ground Control Stations (GCS). Designing effective routing protocols for relaying crucial data in UAV networks is challenging due to dynamic topology, rapid mobility, and limited UAV resources. This paper presents a novel energy-constrained routing mechanism that ensures connectivity, inter-UAV collision avoidance, and network restoration post-UAV fragmentation while adapting without a predefined UAV path. The proposed method employs improved Q learning to optimize the next-hop node selection. Considering these factors, the paper proposes a novel, Improved Q-learning-based Multi-hop Routing (IQMR) protocol. Simulation results validate IQMR's adaptability to changing system conditions and superiority over QMR, QTAR, and Q-FANET in energy efficiency and data throughput. IQMR achieves energy consumption efficiency improvements of 32.27\%, 36.35\%, and 36.35\% over QMR, Q-FANET, and QTAR, along with significantly higher data throughput enhancements of 53.3\%, 80.35\%, and 93.36\% over Q-FANET, QMR, and QTAR.
	\end{abstract}
	
	\begin{IEEEkeywords}
		Unmanned Aerial Vehicle, coverage, collision, netwrok fragmentation,  multi-hop routing, Q-learning.
	\end{IEEEkeywords}

	%
	\IEEEpeerreviewmaketitle

	\section{Introduction}
	
	\IEEEPARstart{N}{atural} disasters like earthquakes have severe consequences, often destroying terrestrial communication networks. A robust and efficient emergency surveillance and communication system is essential to aid search and rescue operations. Unmanned Aerial Vehicles (UAVs) offer great potential in disaster response due to their low cost, high mobility, flexible deployment, and dynamic path optimization characteristics \cite{Mozaffari}\cite{HelpSky}. UAVs with high-resolution cameras and sensors capture real-time data from disaster-stricken areas, providing information about damage, hazards, and locations of victims in need, enabling prompt response efforts \cite{surveillance}. UAVs also serve as communication relays facilitating information exchange between responders and victims and can deliver essential supplies to inaccessible areas \cite{paradigm}\cite{UAVdelivery}\cite{relaying}.
	
	However, ensuring reliable communication among UAVs is challenging due to dynamic topology with disappearing nodes and unstable links \cite{Gupta}\cite{mavanet}. In this context, designing a reliable data routing policy becomes crucial. Effective routing ensures that the information collected by UAVs during the operation can be transmitted reliably to the ground station to enable real-time coordination, situational awareness, and decision-making, facilitating the successful execution of rescue missions \cite{disaster-area}\cite{publicsafety}. 
	
	Achieving reliable data transmission in UAV-assisted emergency communication entails considering various important factors. These include determining the availability of a viable next-hop forwarder within the coverage range, assessing the likelihood of UAV collisions, managing transmission range limitations through multi-hop data transfer, ensuring energy-efficient communication, and addressing network fragmentation challenges. Existing research efforts have not sufficiently integrated these considerations to enable dependable UAV-based data transmission \cite{LAROD}\cite{LCAD}\cite{OLSR}\cite{LADTR}. Therefore, there is a need for an integrated approach that tackles the challenges comprehensively.
	
	This paper addresses the data routing problem in a UAV-assisted emergency communication system for disaster surveillance, where multiple UAVs and a Ground Control Station (GCS) communicate. The UAVs have a Global Positioning System (GPS), cameras, sensors, and wireless communication interfaces to capture and relay information from the surveillance area to the GCS. The paper proposes a novel, Improved Q-learning-based Multi-hop Routing (IQMR) protocol to enable data transmission.  
	
	The contributions of this paper are summarized as follows:
	\begin{enumerate}
		\item We present a novel multi-hop emergency data transmission approach in a dynamic network of interconnected UAVs. This innovative method considers energy management, coverage optimization, and collision prevention while dynamically maneuvering without relying on predetermined UAV paths. Furthermore, our method ensures network continuity in the face of UAV fragmentation. 	Subsequently, we employ an improved Q-learning strategy to determine the optimal next-hop node, further elevating decision-making precision. Collectively, these considerations culminate in a novel data transmission approach named the Improved Q-learning-based Multi-hop Routing (IQMR) protocol.
		\item We define states based on UAV parameters, including residual energy, coverage probability, collision probability, and packet reception status. Residual energy reflects remaining UAV energy levels. Coverage probability assesses signal reception likelihood. Collision probability indicates inter-UAV collision chances. Packet reception status indicates link reliability and includes transmitted packets and acknowledgements (\textit{ACKs})at Layers 2 and 3. Layer 2 handles local communication, while Layer 3 manages end-to-end routing. Layer 2 \textit{ACKs} ensures inter-node packet delivery, whereas Layer 3 \textit{ACKs} guarantees packet transfer across the multi-node path (source UAV to GCS).
		\item We define UAV operational modes, including neighbor discovery, data receiving and transmission, and battery charging.
		\item  We present a novel approach to tackle network fragmentation challenges, ensuring uninterrupted and reliable data transmission even in scenarios with potential network partitioning.
		\item We incorporate dynamic learning rate and discount factor into the protocol to enhance adaptability to varying network conditions. 
		\item Extensive simulations analyze the proposed IQMR's impact on the cumulative system reward for varying parameters, including residual energy, coverage and collision probabilities, packet reception status, and network fragmentation. Comparative evaluations with existing protocols (QMR\cite{QMR}, Q-FANET\cite{QFANET}, QTAR\cite{QTAR}) showcase IQMR's superior performance. Specifically, IQMR achieves energy consumption efficiency improvements of 32.27\%, 36.35\%, and 36.35\% over QMR, Q-FANET, and QTAR, along with significantly higher data throughput enhancements of 53.3\%, 80.35\%, and 93.36\% over Q-FANET, QMR, and QTAR.
	\end{enumerate}

	The remainder of the paper is organized as follows. Section II details related work on the QL-based protocols proposed for UAV data routing and their contributions. Section III describes the system model of the UAV-enabled cellular communication system for disaster surveillance comprising the descriptions of the multi-UAV network, the Gauss-Markov model for UAV mobility, and the communication channel model. Section IV details the novel Improved Q- Learning based Multi-hop Routing  (IQMR) protocol and outlines the three modules of IQMR, namely (i) neighbour discovery module, (ii) estimation module for energy-efficient reliable communication, and (iii) routing decision module. Through extensive simulation, Section V analyses the performance of the proposed IQMR protocol for changing system parameters and compares IQMR with the existing methods. Finally, the paper is concluded in section VI.

	\section{Related Work}

	UAV-assisted communication and networking has attracted much interest from both academia and industry. UAVs are the best candidates for search and rescue operations and disaster management as they can effectively monitor disaster-prone areas inaccessible to humans. However, to provide timely disaster warnings and expedite the rescue process, we need a robust data forwarding mechanism to get the surveillance information to the GCS as quickly as possible. This section provides a detailed survey of the UAV routing protocols and their applicability to harsh environments. 
	
	In addition to the requirements in conventional wireless networks, such as allowing the network to scale, routing data, meeting latency constraints, and maintaining connectivity and required QoS; UAV networks should also consider rapidly changing dynamic topology, disappearing nodes, intermittent links, location awareness and stringent energy constraints. The post-disaster UAV networks may be under appalling circumstances, thus requiring routing protocols to tolerate network dynamics and deliver information efficiently to the ground station without much packet loss, message overhead, and delay. However, the existing routing methods lack flexibility and are limited by space and time. Therefore, there is a need for a tailor-made routing approach that suits the application-specific requirements of aerial networks.
	
	Several works have investigated the applicability of existing routing schemes of MANETs and VANETs, like topology-aware and position-based protocols, for possible use in UAV networks. The topology-aware routing method includes static, proactive, reactive, and hybrid routing protocols. Static routing protocols have fixed routing tables computed and loaded to the UAVs ahead of the flight and remain unchanged during the operation. Data Centric Routing (DCR) for one-to-many transmission and Load Carry And Delivery (LCAD)  for improved throughput are explored for UAV networks. However, the static routing protocols are not dynamic and fault tolerant \cite{DCR}\cite{LCAD}.
	
	On the other hand, the proactive protocol updates the routing table regularly. Optimized Link State Routing (OLSR) protocol is the most studied multi-point relay forwarding proactive method for UAVs \cite{OLSR}. Better Approach to Mobile Ad-hoc Networking (BATMAN) protocol is reviewed as an improvement to the OLSR \cite{BATMAN}. It exhibits faster convergence as it does not calculate the complete route but finds the next best hop for each destination node. However, the routing table updation mandates exchanging a good number of node control messages, creating network overhead. Contrarily, reactive protocols support on-demand routing to overcome the message overhead problem of proactive protocols supporting source and hop-by-hop routing like Ad-hoc On-Demand Distance Vector (AODV) and Dynamic Source Routing (DSR) for use in airborne networks \cite{AODV-DSR}. However, their lengthy route discovery process increases the network latency. Hence proactive and reactive protocols are not appropriate for transmitting data in highly dynamic UAV networks. The hybrid protocol intends to overcome the drawbacks of the message overhead in proactive and route discovery delay in reactive approaches. They are best suitable for large-scale networks with several sub-network areas where intra-zonal and inter-zonal routing employ proactive and reactive routing, respectively, as in the Zone Routing Protocol (ZRP) \cite{ZRP}. However, the dynamicity of the UAV networks makes it difficult for the protocol to access and manage the information.
	
	The position-based protocol uses the UAVs' geographical location information for either a single path (interaction between UAVs and fixed nodes) or multi-path data forwarding. The position-based forwarding method includes Delay Tolerant Network (DTN) and Non-Delay Tolerant Network (NDTN) routing. DTN routing methods are primarily designed for harsh environments, for instance, in the aftermath of disasters, where regular communication breaks down with intermittent end-to-end connectivity and network getting partitioned for long durations. In such scenarios, the routing protocols deployed for infrastructure-based internet will not be able to handle data transmission as multi-step request response, acknowledgments, and timed-out transmissions may only sometimes work due to long delays. The protocol architecture incorporates a store-carry-forward method to achieve delay tolerance, in which the mobile node stores and carries the data till it can replicate in one or more nearby nodes. The most used DTN protocols are Location-Aware Routing for Opportunistic Delay Tolerance (LAROD), Location-Aided Delay Tolerant Routing (LADTR) and Jamming-Resilient Multi-path Routing (JARMROUT) \cite{LAROD}\cite{LADTR}\cite{JARMROUT}. However, these protocols require an adequate buffering capacity to store the data until it can forward it.
	
	Conversely, NDTN protocols are studied for applications requiring minimum data transmission delay. NDTN routing methods like Geographical Location Sharing Routing (GLSR) and Reactive Greedy Reactive (RGR) protocol are extensively deployed in high-density UAV networks \cite{GLSR}\cite{RGR}. However, the network performance in terms of packet delivery rate and latency may not be minimal in the case of a highly dynamic network topology where data delivery may fail, and a recovery approach may be required.
	
	The complex and diverse flight environments have caused airborne networks to be in an unpredictable random state. Therefore the traditional routing protocols will not be able to adapt to changing networks in real-time and cater to the needs of the UAV networks. Hence flexible and autonomous machine learning-based protocols have been envisioned as promising solutions complaisant with the unique requirements of UAV networks. Reinforcement learning is an adaptive learning method belonging to the machine learning category. The proposed work implements the Q-learning-based reinforcement learning technique for multi-hop data transfer. The following parts of this section details the various Q-learning based routing techniques for UAV communication.

Q-Learning-Based Geographic (QGeo)\cite{QGeo} protocol incorporates packet travel speed to determine the next forwarder based on packet travel time, MAC delay, and transmission delay. Reward Function Learning for QL-Based Geographic Routing Protocol (RFLQGEO)\cite{RFLQGEO} extends QGeo by considering UAV position information obtained through GPS to compute the distance progress of data towards the sink while considering MAC and transmission delays. Q-Learning-Based Cross-Layer Routing Protocol (QLCLRP)\cite{QLCLRP} integrates carrier sense multiple access with multi-channel synchronization to ensure reliable data transmission in UAV networks. Q-Learning-Based Balanced Path Routing (QLBR) \cite{QLBR} is a routing protocol designed to achieve load balancing and efficient resource utilization by dynamically selecting balanced paths based on network congestion, link quality, and available resources. Q-Learning-based Multi-Objective Routing (QMR)\cite{QMR} focuses on optimizing delay and energy consumption in data routing. Q-Learning-based Routing Protocol for FANET  (Q-FANET) \cite{QFANET} aims to enhance the Quality of Service in dynamic UAV networks by minimizing delay and considering channel conditions and residual energy levels. QL-Based Topology-Aware Routing (QTAR) \cite{QTAR} leverages two-hop neighbor information to facilitate efficient data forwarding, considering factors like transmission delay and changing network conditions. However, QGeo, RFLQGEO, QLCLRP, QLBR, and Q-FANET overlook energy efficiency as a routing parameter and lack adaptability to changing network characteristics such as topology, link quality, mobility, and obstacles. Although QMR estimates the link quality using the expected transmission count to update the Q-value, considering the SIR  as a metric would provide a more accurate estimation of the link status. In QTAR, the maintenance of two-hop neighbor information leads to increased network overhead.

Hence, there is a requirement for specialized protocols that address the unique needs of airborne networks, offering adaptability to changing topology and demand-specific routing, particularly in scenarios without predetermined UAV path planning. QL-based adaptive routing protocols have been explored, and they often lack simultaneous multi-objective optimization of parameters crucial for data routing in dynamic UAV networks. Our research proposes an integrated approach to develop a novel packet forwarding method that incorporates factors such as UAV energy consumption, coverage and collision constraints, and packet reception status without prior trajectory planning. Additionally, our approach dynamically adjusts QL parameters (learning rate and discount factor) and neighbour discovery (incidence of \textit{HelloInterval}) to respond effectively to changing network characteristics. To the best of our knowledge, this proposed novel approach has not been addressed in the existing literature.

	\begin{table*}	
		\tiny
		\centering
		\caption{Comparison of Q-learning based protocols}
		\resizebox{\textwidth}{!}{
				\begin{tabular}{|p{1cm}|p{3cm}|p{1cm}|p{1cm}|p{1cm}|p{1cm}|p{1cm}|}
						\hline
						\multirow{1}{*}{\textbf{Protocol}}&\centering {\textbf{Approach}}&\multicolumn{5}{c|}{\textbf{Parameter}}\\
						\cline{3-7}
						&{}&End-to-end delay&Energy consumption&Adaptive Q-learning parameters&Adaptive \textit{HelloInterval}&Mobility control\\
						\hline
						QGeo\cite{QGeo}&Utilizes MAC delay, link error, and localization error to determine the link quality and avoid the local minimum problem.&\centering \ding{51} &\centering \ding{55}& \centering \ding{55}& \centering \ding{55}&Random\\
						\hline
						RFLQGEO\cite{RFLQGEO}&Utilizes MAC delay, link error, localization error, residual energy and distance progress toward destination to determine the forwarding node.&\centering \ding{51} &\centering \ding{51}& \centering \ding{55}& \centering \ding{55}&Random\\
						\hline
						QLCLRP\cite{QLCLRP}&Adopts cross-layer adaptive QL-based routing protocol to minimize delay, link error to achieve reliable data transfer.&\centering \ding{51} &\centering \ding{51}& \centering \ding{55}& \centering \ding{55}&Random\\
						\hline
						QLAR\cite{QLAR}&Utilizes the mobility level of each node to adapt to static and dynamic routing in response to topology changes.&\centering \ding{55} &\centering \ding{55}& \centering \ding{55}& \centering \ding{55}&Random\\
						\hline
						QMR\cite{QMR}&Jointly minimizes the delay and energy by designing a multi-objective reward function adaptive to changing network conditions.&\centering \ding{51} &\centering \ding{51}& \centering \ding{51}& \centering \ding{55}&Random\\
						\hline
						Q-FANET\cite{QFANET}& Minimize jitter, end-to-end delay, and latency in highly dynamic conditions  with local minimum avoidance.&\centering \ding{51} &\centering \ding{55}& \centering \ding{51}& \centering \ding{55}&Random\\
						\hline
						QTAR\cite{QTAR}&Utilizes two hop neighbour information and adopts a multi-objective reward function to optimize delay, energy consumption, to avoid routing loop.&\centering \ding{51} &\centering \ding{51}& \centering \ding{51}& \centering \ding{51}&3D Gauss Markov\\
						\hline
					\end{tabular}}
		\label{tab:multicol}
	\end{table*}	
	
	
	Therefore, there still exists a need for protocols that cater to specific requirements of airborne networks that adapt to changing topology and demand-specific routing, especially for scenarios without prior UAV path planning. This section discusses UAV-related routing protocols and highlights that the existing packet forwarding methods, including topology-aware and position-based protocols, are not inherently suitable for aerial networks due to rapidly changing topology, high mobility, intermittent connectivity, and resource constraints. Though adaptive routing protocols based on Q-learning are investigated, they do not consider simultaneous multi-objective optimization of parameters influencing the data routing in dynamic UAV networks. Our work focuses on an integrated approach to designing a new packet forwarding method without prior trajectory planning considering parameters including UAV energy consumption, coverage and collision constraints and packet reception status. On the other hand, the proposed work adaptively adjusts the Q-learning parameters (learning rate and discount factor) and \textit{ HelloInterval} according to the changing network characteristics. To the best of our knowledge, previous works in literature do not address our proposed approach.

	\section{System Description}
	\subsection{Network Model}
	The paper considers a network of low-altitude UAVs that communicate with each other and a Ground Control Station. The network consists of $M$ UAV nodes denoted as $\mathbb{U} = [U_1,\dots, U_M]$. Each UAV, represented by $U_i$, is characterized by parameters such as residual energy ($E^{res}$), packet reception status ($P^{rs}$), coverage ($P^{cov}$), and collision probabilities ($P^{coll}$). The GCS has a fixed location within the network, while the UAVs are randomly distributed in a three-dimensional (3D) space.
	
	The UAVs, stationed initially on the ground at the start location, ascend toward the designated base location in the airspace. The start and base locations fall in 3D space of cylindrical geometry with radius R and maximum height H above the ground plane. The UAVs' start location is at the center of the cylindrical region  $(x_0, y_0, h_0)$ on the ground. In contrast, the base locations $(x_i^b, y_i^b, h_i^b)$ are randomly distributed within the 3D cylindrical space as seen in Fig. 1. All UAVs are confined within a 3D space throughout the operation and move in discrete time slots using the Gauss Markov Mobility Model (GMMM) without prior trajectory planning.
	
	\begin{figure}
		\centering
		\includegraphics[width=6cm, height=6cm]{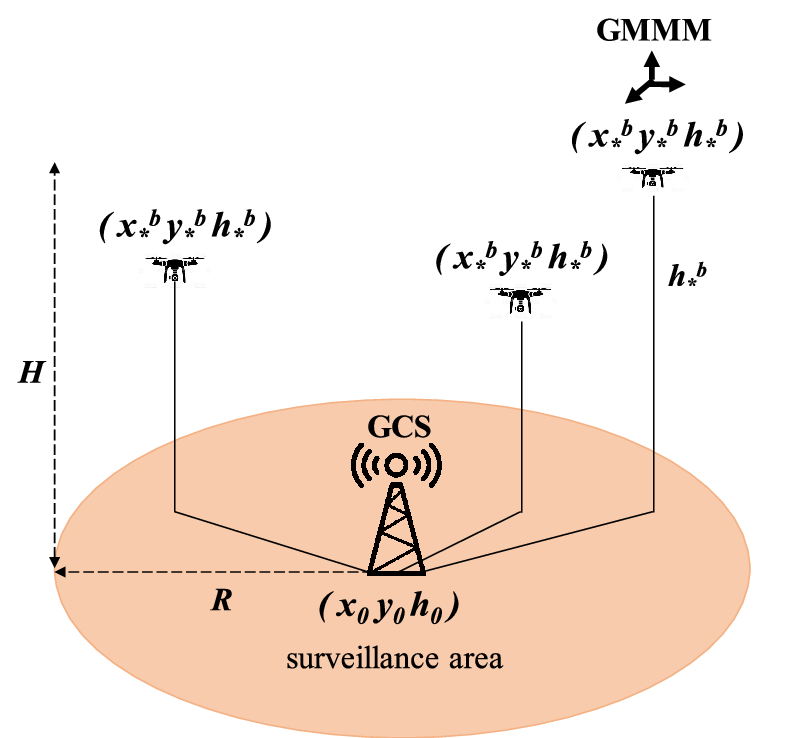}
		\caption{Multiple UAV network with 3D GMMM. $(x_*^b,y_*^b,h_*^b)  ($* = [1\dots M]$)$ representing the base location of UAVs in air, $(x_0,y_0,h_0)$  indicating location of GCS and UAVs start point.}
	\end{figure}
	
	\subsection{Gauss Markov Mobility Model}
	The UAVs follow the 3D Gauss-Markov Mobility Model to guide their movement process in the air \cite{LADTR}. Initially, the UAVs fly to their respective base locations in the air from the start point on the ground. From the base location onward, the UAVs take random positions in the network independently following the GMMM. Each UAV is assigned a mean speed ($s$), direction ($d$), and vertical pitch ($p$) with respect to the ground plane. Vertical pitch refers to the angle of ascent or descent affecting the UAV's altitude change. The UAV calculates $s$, $d$ and $p$ at every set time $(n)$ using the Gauss-Markov equations as 
	\begin{equation}
		s_n = \alpha s_{n-1}+(1-\alpha) \bar{s} + \sqrt{(1-\alpha^2)}s_{x_{n-1}}
	\end{equation}
	\begin{equation}
		d_n = \alpha d_{n-1}+(1-\alpha) \bar{d} + \sqrt{(1-\alpha^2)}d_{x_{n-1}}
	\end{equation}
	\begin{equation}
		p_n = \alpha p_{n-1}+(1-\alpha) \bar{p} + \sqrt{(1-\alpha^2)}p_{x_{n-1}}
	\end{equation}
	
	The tuning parameter $\alpha$ takes a value within the range of (0,1). The mean speed, direction, and pitch are denoted as $\bar{s}$, $\bar{d}$, and $\bar{p}$, respectively. To introduce randomness to the new speed, direction, and pitch, random variables $s_{x_{n-1}}$, $d_{x_{n-1}}$, and $p_{x_{n-1}}$ are chosen from uniform distributions $[s_{min}, s_{max}]$, $[d_{min}, d_{max}]$, and $[p_{min}, p_{max}]$, respectively. The next position of the UAV ($x_{n+1}, y_{n+1}, h_{n+1}$) is determined from the current location ($x_n, y_n h_n$), the speed ($s_n$), direction ($d_n$), and pitch ($p_n$) as
	\begin{equation}
		x_{n+1} = x_{n}+s_{n}cos(d_{n})cos(p_{n})
	\end{equation}
	\begin{equation}
		y_{n+1} = y_{n}+s_{n}sin(d_{n})cos(p_{n})
	\end{equation}
	\begin{equation}
		h_{n+1} = h_{n}+s_{n}sin(p_{n})
	\end{equation}
	
	On reaching the location, the UAVs stay for a pause time $(t_p)$ chosen from a uniform random distribution $[t_{min}, t_{max}]$. During $t_p$, the UAVs relay the disaster-related information to GCS via a multi-hop UAV path.
	\subsection{Channel Model}
	The propagation through the wireless channel is a combination of distance-dependent path loss attenuation and small-scale fading. We adopt the standard power-law path loss model and define the path loss function as $l(r_{ij}, h_i,\alpha )= (r_{ij}^2 + h_i^2)^{-\alpha/2}$ where $r_{ij}$ denotes the horizontal distance between the $i^{th}$ transmitting UAV and $j^{th}$ receiving UAV, $h_i$ is the operating height of the $i^{th}$ UAV and $\alpha$ is the path loss exponent \cite{pathloss}. We assume Nakagami-$m$ small-scale fading channel with fading parameter $m$ computed as $m \triangleq 2(K+1)/(2K+1)$ where $K$ is the Rician $K$-factor characterizing the ratio between the powers of direct and scattered paths \cite{nakagami}. Therefore, the channel power fading gain $g_{ij}$ follows the Gamma distribution $\Gamma(m,\frac{1}{m})$.
	
	Assuming interference-limited scenario \cite{Ilimited}, we define the Signal to Interference Ratio (SIR) at $j^{th}$ receiving UAV from $i^{th}$ transmitting UAV as
	\begin{equation}
		SIR_{ij} \triangleq \frac{g_{ij}l(r_{ij},h_i)}{I}
	\end{equation}
	$I$ is the aggregate interference power defined as $ I \triangleq \sum_{x\in\mathbb{U}\setminus\{i\}}g_{xj}l(r_{xj},h_x)$
	
	\subsection{Problem Formulation}
	The UAVs capture the data during surveillance and transmit the acquired information to the GCS. The objective is to maximize the residual energy, packet reception status, and coverage probability while minimizing the chances of collision. The multi-hop data routing problem formulation for energy-efficient, reliable communication over the considered distributed UAV network is as follows.
	\begin{equation}
	max(\mathcal{E}_n^{res}(i)), max(\mathcal{P}_n^{rs}(i)), max(\mathcal{P}_n^{cov}(i)), min(\mathcal{P}_n^{coll}(i))
\end{equation}
subject to:
\begin{equation}
	\mathcal{E}_n^{res}(i) > \delta_{th} 
\end{equation}

\begin{equation}
	\mathcal{P}_n^{cov}(i)=\mathbb{E}[SIR_n(i)\geq\theta_{th}] 
\end{equation}
\begin{equation}
	\mathcal{P}_n^{coll}(i) < \phi_{th}	\hspace{0.2cm}
\end{equation}
\begin{equation*}
	\forall n \in N, \forall i \in M
\end{equation*}

The objective function in (8) considers the normalized parameters. The normalization of the parameters follows the below relation, 
\begin{equation}
	\hat{x}=\frac{x_{actual}-x_{minimum}}{x_{maximum}-x_{minimum}}
\end{equation}
In (12), $x$ indicates the parameter to be normalized. The network runs for $N$ number of episodes. The equations from (9) to (11) state the constraints for the  objective (8). Under condition (9), successful data transmission between UAVs is feasible when their remaining energy levels exceed the defined threshold ($\delta_{th}$). Constraint (10) stipulates the reliability criteria for connectivity, demanding a signal-to-interference ratio higher than or equal to the predefined threshold ($\theta_{th}$). Constraint (11) dictates collision prevention, requiring UAVs to relocate randomly if collision probability exceeds a specific threshold ($\phi_{th}$) to avert physical contact.The Q learning a reinforcement learning method is adopted to optimize the parameters in (8) and dynamically learn the most suitable routing mechanism for data transfer between the UAVs and GCS. The paper's next section enumerates the solution for the optimization problem in (8).

\section{Improved Q-learning-based Multi-hop Routing Protocol for Energy Efficient Reliable Communication}
This section proposes an Improved Q-learning-based Multi-hop Routing (IQMR) protocol for improved energy efficiency and throughput. Fig. 2 enumerates the block diagram of the proposed IQMR protocol, which includes three modules: (i) neighbour discovery module, (ii) estimation module for energy-efficient reliable communication, and (iii) routing decision module.
\begin{figure}
	\centering
	\includegraphics[width=7cm, height=6cm]{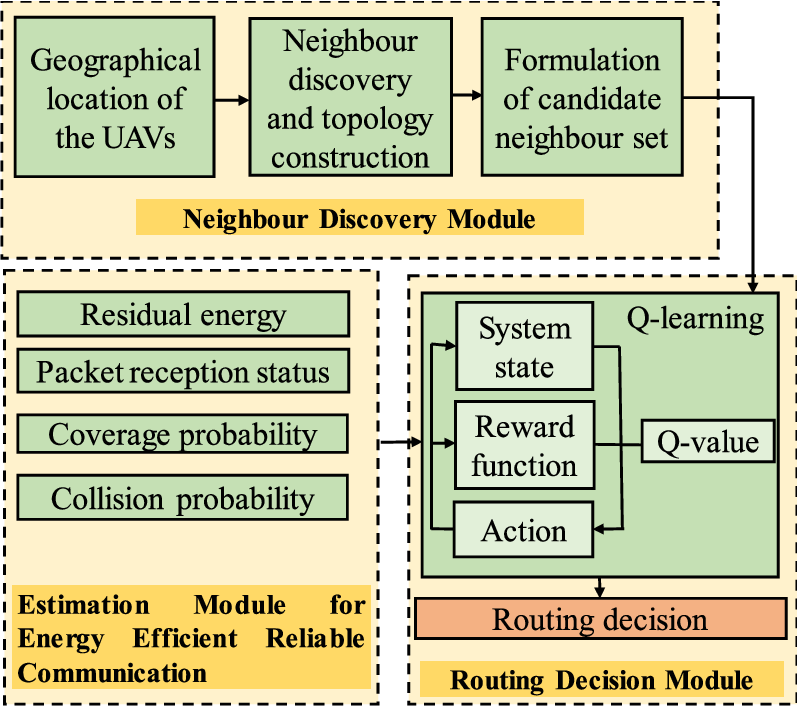}
	\caption{Framework of Improved Q-learning-based Multi-hop Routing Protocol for Energy Efficient Reliable Communication}
\end{figure}

\subsection{Neighbour Discovery Module}
The neighbour discovery module, described in \textit{Algorithm 1},  constructs and updates the network topology by maintaining a  neighbour table ($N_T$). Neighbour discovery involves exchanging \textit{HelloMessage} over every \textit{HelloInterval}. Each \textit{HelloMessage} contains information such as UAV location, residual energy, packet reception status, learning rates, discount factor, and Q-value. The receipt of the \textit{HelloMessage} establishes the UAV's neighbour set, consisting of the nodes from which the UAV receives the message. A subset of neighbour UAVs  limited to a specific sector $\mathbb{S}$ are considered as candidate neighbours. The sector $\mathbb{S}$ of radius  $R_t$ (uniform UAV transmission range) and spans an angle of $\pi$  around the S-D axis (source node and destination node (GCS) axis). The selection of candidate neighbours is to identify UAVs that have the potential to move the data closer to the destination node, as illustrated in Fig.3 \cite{my}. The neighbour table stores information in the \textit{HelloMessage} received from the candidate neighbours.

Determining the appropriate timing for the \textit{HelloInterval} is crucial to adapt to the dynamically changing network. The \textit{HelloInterval} should be set to a frequency allowing timely updates of the neighbour table while minimizing unnecessary overhead. By evaluating the Link Sustenance Time ($t^{LST}$), we can estimate the incidence of \textit{HelloInterval} based on the acquired topology information. The $t^{LST}$  represents the duration the link between communicating UAVs remains stable and reliable, ensuring collision free uninterrupted connectivity.

The estimation of $t^{LST}$ is influenced by the relative motion between the two communicating UAVs. The three scenarios for determining $t^{LST}$ include UAVs (i) receding, (ii) approaching, and  (iii)  moving equidistant apart. Let $U_i$ and $U_j$ be the two communicating UAVs, and $R_t$ be the uniform transmission range of each UAV. The relative distance ($d_{ij}(n)$) between the two UAVs $U_i$ and $U_j$ at $n^{th}$ discrete time instant is determined using the Euclidean distance formula as
\begin{multline}
	d_{ij}(n) =\\ \sqrt{[x_i(n)-x_j(n)]^2 + [y_i(n)-y_j(n)]^2 + [z_i(n)-z_j(n)]^2}	
\end{multline}
Assume UAVs $U_i$ and $U_j$ moving at speeds $s_i$ and $s_j$, respectively.  When the UAVs are receding, the expected time for their separation to exceed their transmission range ($R_t$) is  $t^{LST}_{ij}(n) = |[R_t - d_{ij}(n)]/[s_i+s_j]|$. Conversely, when UAVs are approaching, and a collision is imminent, $t^{LST}_{ij}(n)= [d_{ij}(n)/|s_i-s_j|]$ represents the time at which their separation becomes zero. Based on these estimations, updating the next occurrence of the  \textit{HelloInterval}  as $H_I(next) = H_I(n) + t^{LST}_{ij}(n)$ allows UAVs to maintain communication within the dynamic network environment while avoiding collision. In the case where UAVs are moving equidistantly apart, calculating $t^{LST}_{ij}(n)$ is unnecessary since their relative positions remain unchanged over time, causing the \textit{HelloInterval} to repeat at each time instant $n$ consistently. Suppose the UAVs fail to receive the \textit{HelloMessage}  associated with an existing record within a specific expiration time of \textit{HelloInterval} (set to $300$ ms), the corresponding entry is removed from the neighbor table.
\begin{figure}
	\centering
	\includegraphics [width=0.75\linewidth] {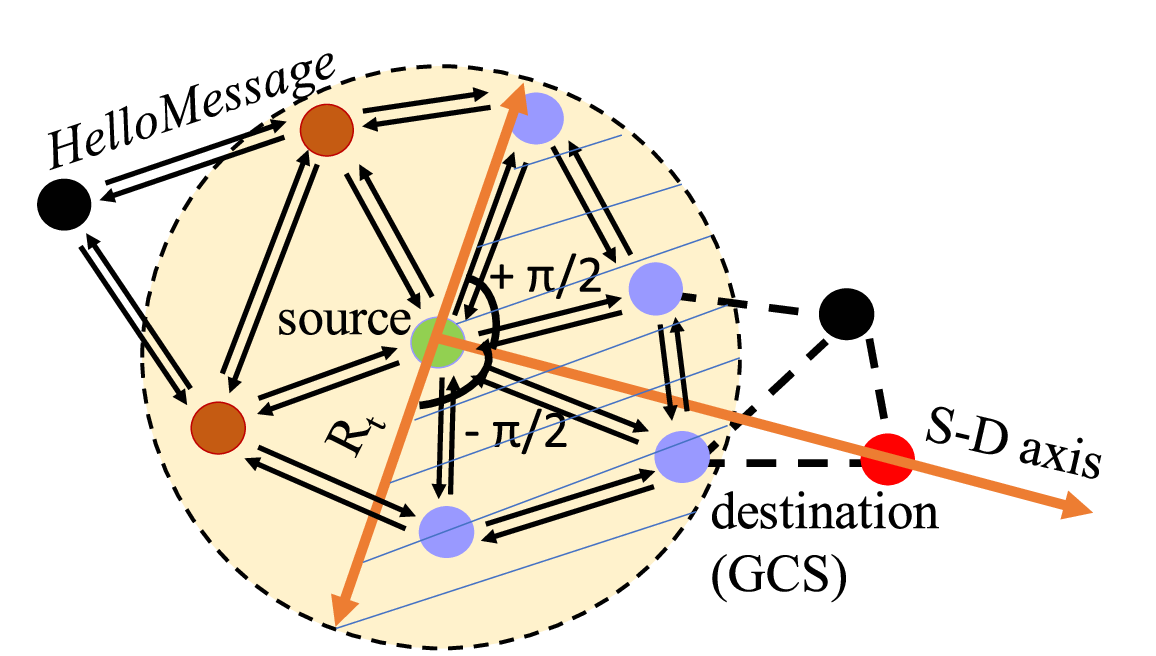}
	\caption{Representation of sector $\mathbb{S}$  (shaded region)  and formulation of candidate neighbour set (nodes within $\mathbb{S}$). Representation of nodes are as black-non neighbours, brown-neighbours, blue-candidate neighbours, green-source, red-destination}
\end{figure}

	\begin{algorithm}
		\caption{: Hello Message Exchange for Topology Construction}\label{algo1}
		\textbf{Input:} \\
		Location of ${M}$ UAVs and GCS, \textit{HelloMessage} and sector $\mathbb{S}(R_t,\pi)$ . \\
		\textbf{Output:} Updated neighbour table ($N_T$) and topology construction. \\
		\textbf{Initialization:}\\
		1.Set network area to $\pi R^2$ m\textsuperscript{2}\\
		2.Set initial energy for UAVs $E_{\text{full}}$\\
		3.Set UAV identification number as $U_i = [U_1,\ldots, U_M] $\\
	
			\raggedright{\textbf{Phase 1: Broadcast \textit{HelloMessage}}}\\
			\begin{algorithmic}[1]
				\FOR{each UAV $U$ in $U_1$ to $U_M$}
				\STATE broadcast \textit{HelloMessage}
				\ENDFOR
			\end{algorithmic}
			\raggedright{\textbf{Phase 2: Neighbour table construction}}\\
			\begin{algorithmic}[1]
				\FOR{each received \textit{HelloMessage}}
				\STATE  $U_i =$ originator node id
				\STATE $Location_{U_i} =$ location of $U_i$
				\IF{$Location_{U_i} \in \mathbb{S}$} 
				\STATE {$N_C=U_i$} // update candidate neighbour set
				\IF{$U_i\in N_T$}
				\STATE update existing record of $U_i$
				\ELSE
				\STATE add a new record for $U_i$
				\ENDIF
				\ENDIF
				\ENDFOR
				\STATE update $H_I$ 
			\end{algorithmic}
		\end{algorithm}
		\subsection{Estimation Model for  Energy Efficient Reliable Communication}
		This section presents estimation models for the proposed IQMR protocol. The model determines residual energy, packet reception status, coverage and collision probabilities. 
		\subsubsection{Residual Energy Estimation Model}
		The proposed work employs a distance-dependent energy consumption model  for estimating the energy expended for data transmission \cite{energy}. Suppose the distance ($r$) between the transmitting UAV and the receiving UAV is less than a certain threshold ($r_0$); we use a free space channel model ($d^2$ power loss); otherwise,  multi-path fading model ($d^4$ power loss). Thus the expression for energy expended to transmit $k$ bits of data over a distance $r$ is \cite{eng-eqs},
		\begin{equation}
			\mathcal{E}^{tx}(k,r)=
			\begin{cases}
				\epsilon_{elec}k + \epsilon_{amp-fs}kr^2 \hspace{1.5cm} r \leq r_0\\
				\epsilon_{elec}k + \epsilon_{amp-mp}kr^4 \hspace{1.5cm} r>r_0
			\end{cases}
		\end{equation}
		where $\epsilon_{elec}$ and $\epsilon_{amp}$ are system constants. $\epsilon_{elec}$ is the energy dissipated by the radio to run the transmitter-to-receiver circuitry (J/bit), $\epsilon_{amp-fs}$ is the distance-dependent amplifier energy with free space channel model (J/bit/m$^2$) and $\epsilon_{amp-mp}$ is the distance-dependent amplifier energy with multi-path channel model (J/bit/m$^4$). 
		
		The energy model for UAV flight is defined as \cite{engfly}
		\begin{equation}
			\mathcal{E}^{fly} = \frac{\epsilon_{payload}w + \epsilon_{hover}\tau}{1-\frac{\epsilon_{payload}}{\epsilon_{density}}\tau}
		\end{equation}
		where $\epsilon_{payload}$ is the power consumed per kilogram of payload (kW), $w$ is the payload weight (kg), $\epsilon_{hover}$ is the power used for hovering (kW), $\epsilon_{density}$ is the battery energy density (kJ/kg), and $\tau$ is the travel time (s). 
		
		The residual energy corresponding to each $i^{th}$ UAV at $n^{th}$ discrete time interval is defined as,
		\begin{equation}
			\mathcal{E}_n^{res}(i )= \mathcal{E}_n^{full}(i) - [\mathcal{E}_n^{tx}(i)+\mathcal{E}_n^{fly}(i)]
		\end{equation}
		where $\mathcal{E}^{full}$ is the full battery capacity of each UAV.
		\subsubsection{Packet Reception Status}
		The overall data throughput in the network is impacted by the packet reception status, which encompasses transmitted data packets at layer 2 ($PAC^{tx}(L2)$) and layer 3 ($PAC^{tx}(L3)$), and acknowledgements at layer 2 ($ACK(L2)$) and layer 3 ($ACK(L3)$). The successful reception of these packets plays a crucial role in ensuring effective data transfer to the Ground Control Station (GCS). The packet reception status during each episode of data transmission is determined by the reception status at Layer 2 and Layer 3, respectively, assuming that certain packets are always transmitted ($PAC^{tx}(L2) \neq 0, PAC^{tx}(L3)\neq 0$).
		\begin{equation}
			\mathcal{P}_n^{rs}(i)=  \sum_{n=0}^{N} \mathcal{P}_n^{rs}(L2)(i) + \mathcal{P}_n^{rs}(L3)(i) \hspace{0.1cm}\forall\hspace{0.1cm} i \in M
		\end{equation}
		\begin{multline}
			\mathcal{P}_n^{rs}(i)= \sum_{n=0}^{N}  \frac{ACK_n(L2)(i)}{PAC^{tx}_n(L2)(i)} + \frac{ACK_n(L3)(i)}{PAC^{tx}_n(L3)(i)} \hspace{0.1cm} \forall\hspace{0.1cm} i \in M
		\end{multline}
		In (18), the expressions $PAC^{tx}_n(L2)(i)$ and $PAC^{tx}_n(L3)(i)$ denote the cumulative number of Layer 2 and Layer 3 packets transmitted, respectively, by the $i^{th}$ UAV up to the $n^{th}$ episode of data transfer. Similarly, $ACK_n(L2)(i)$ and $ACK_n(L3)(i)$ represent the cumulative count of Layer 2 and Layer 3 acknowledgements received by the $i^{th}$ UAV corresponding to the packets it has transmitted. These parameters track the ongoing data transfer process.
		\subsubsection{Coverage Probability}
		The coverage probability represents the minimum SIR requirement between communicating UAVs for reliable data transfer. Following the analytical model in \cite{my} $\mathcal{P}^{cov}_n(i)$ for $i^{th}$ UAV at discrete time $n$ is expressed as follows.
		\begin{equation}
			\mathcal{P}^{cov}_n(i) = \mathbb{E}[SIR_n(i)\geq\theta_{th}]
		\end{equation}
		where $\theta_{th}$ is the minimum SIR requirement.
		\subsubsection{Collision Probability}
		The collision probability indicates the likelihood of two UAVs colliding when their separation is below the minimum required threshold ($r_{min}$) to avoid a collision. Based on the mathematical model in \cite{my}, the expression for the collision probability $\mathcal{P}^{coll}_n(i)$ of the $i^{th}$ UAV at the $n^{th}$ discrete time instant can be formulated as follows.
		\begin{equation}
			\mathcal{P}^{coll}_n(i) = 1- exp\bigg(\frac{-r^2_n(i)}{2 \xi^x_n(i) \xi^y_n(i)}\bigg)  < \phi_{th}
		\end{equation}
		where $r_n$ is the distance between the UAVs and $\xi^x_n$ and $\xi^y_n$ are divergence in the UAV trajectory in $x$ and $y$ directions respectively at $n^{th}$ instant of data transfer and $\phi_{th}$ is the minimum collision probability requirement. 
		
		\subsection{Routing Decision Module}
		This section proposes a Routing Decision Module in the UAV-assisted emergency communication system that employs  Q-learning, a reinforcement learning-based framework, to determine the appropriate next-hop node for transmitting surveillance data to the GCS through a multi-hop path. The module leverages information from the Neighbour Discovery Module and Estimation Module, the operational states and modes of both the source UAV and neighbouring UAVs, and associated reward values. 
		
		\subsubsection{Reinforcement Learning based Multi-hop Routing}
		The next-hop node selection depends on the UAVs’ operating mode and current state. The operating modes are neighbour discovery, receive, transmit, and charge, discussed later in this section. The state of the UAVs includes residual energy, packet reception status, coverage and collision probabilities. Based on these parameters an approximate action towards data forwarding, which leads to maximized immediate reward value, is chosen. Over time the UAVs learn the best next-hop node to forward the data that maximizes the cumulative reward values between the source UAV and the GCS. For reinforcement learning, we define a tuple $(S, A, R)$ where $S$ represents the state space, $A$ indicates the action space and $R$ specifies the reward space. Below is a detailed description of the state, action, and reward space.
		\paragraph{States}
		At $n^{th}$ discrete time interval, the network state is given as $	\boldsymbol{S}_n = [\boldsymbol{E}_n^{res}(i), \boldsymbol{P}_n^{rs}(i), \boldsymbol{P}_n^{cov}(i), \boldsymbol{P}_n^{coll}(i) ]$ The vectors $\boldsymbol{E}_n^{res}(i)$, $\boldsymbol{P}_n^{rs}(i)$, $\boldsymbol{P}_n^{cov}(i)$ and $\boldsymbol{P}_n^{coll}(i)$ represent the residual energy, packet reception status, coverage, and collision probabilities in their normalized form for a set of $\mathbb {U} $ UAVs.
	
		\paragraph{Actions}
		The selection of the next-hop node incorporates normalized system parameter values, enabling routing decisions according to the dynamic conditions of the network. The action space, denoted as $A_n$, is defined as $A_n = a_n(i)$, where $a_n(i)$ represents the routing decision of the $i^{th}$ UAV at $n^{th}$ discrete time. 
		\paragraph{Rewards}
		The reward function maps states to actions by assigning positive reward values corresponding to the action taken. These rewards define the effectiveness of successful data transmission in a dynamic network. Thus the reward function corresponding  to the action is defined as a joint metric considering the normalized values of the parameters in the objective function (8) as,
		
		\begin{equation}
			R(s_n,a_n)=
			\begin{cases}
				w_1(1-\mathcal{P}_n^{coll})+ w_2 \mathcal{P}_n^{rs}(L3)+\\ w_3\mathcal{P}_n^{rs}(L2)+ w_4\mathcal{P}_n^{cov}+ w_5\mathcal{E}_n^{res}  , & \text{if $N_H \neq 0$ }\\
				0, & \text{otherwise}
			\end{cases} 
		\end{equation}
		where $w_1 >> w_2 >> w_3 >> w_4 >> w_5$ and $N_H$ is the number of candidate neighbour nodes.
		
		The weights $w_1, w_2, w_3, w_4$, and $w_5$ are chosen based on the significance of the corresponding parameter on data routing. The collision of two UAVs can disrupt the working of the entire network by changing the topology, causing irreparable damage to the UAVs and leading to the loss of information. Accordingly, the $P^{coll}$ receives the highest weightage among other system parameters. $P^{rs}$ receives the next highest weightage as it directly reflects the reliability of data transmission. Successful data transfer between the source UAV and the GCS is of greater significance than between the UAVs since the ultimate goal is to ensure that the data reaches the GCS for further processing and analysis. As a result, the parameter $P^{rs}(L3)$, which represents the success of data transfer between the source and the GCS, is assigned a higher weightage than $P^{rs}(L2)$, which represents the success of data transfer between UAVs. The  $P^{cov}$ provides information about the network characteristics, such as the topology and wireless channel conditions. While $P^{cov}$  indicates the reliability of the link between communicating nodes, $P^{rs}$ provides real-time feedback on the actual reception of transmitted packets. Thus $P^{rs}$ receives higher weightage than $P^{cov}$. On the other hand, the $E^{res}$ represents the remaining battery capacity of the UAVs. While the energy level is an important parameter for UAV operations, its weightage in the routing decision process is lower than $P^{cov}$  because ensuring a robust communication link precedes energy considerations. Without a  stable link, even with sufficient energy, data transmission may be prone to disruptions and failures. Therefore, to prioritize the reliability of the communication link, the weightage assigned to $P^{cov}$  is higher than that given to $E^{res}$. A minimum reward of zero for network fragmentation ($N_H=0$) serves as a discouragement to decrease the frequency of such occurrences.

		The solution to the formulated reinforcement learning problem is policy $\pi$, which maps $\pi: \boldsymbol{S} \rightarrow \boldsymbol{A}$ with $\pi(s)$ being the optimal policy taken in state $s\in \boldsymbol{S}$. An optimal policy $\pi^*$  maximizes the expected discounted reward as,
		\begin{equation}
			\pi^* = arg \max_{\pi}\mathbb{E}\bigg[\sum_{n=0}^{\infty}\gamma_nR(s_n,\pi(s_n))\bigg]
		\end{equation}
		The discount factor,  $\gamma$, is crucial in balancing immediate rewards and long-term benefits. The algorithm accommodates uncertainties and ensures convergence towards an optimal routing policy by adjusting  $\gamma$ within the range   $(0,1)$.
	
		\subsubsection{Improved Q-learning ($Q(\lambda$)) Framework for Multi-hop Routing}
		Designing routing protocols using a model-based approach in UAV networks becomes impractical as rewards can only be determined when the destination node receives data packets and the source node receives acknowledgements. Hence, a value-based model-free reinforcement learning method like Q-learning is necessary. 
		The improved Q-learning scheme incorporates eligibility traces to assign rewards to previous states, capturing temporal dependencies and enhancing learning efficiency. This approach allows entities to continually adapt their actions according to the reward function, maximizing their outcomes. Each element of the Q-matrix is a function of the UAV's state and action and is updated at discrete time instants using the below relation,
		\begin{multline}
			Q^{new}(s_n,a_n) \leftarrow Q^{old}(s_n,a_n) + \beta \big[R_{n+1} +\\ \gamma \max_{a}Q(s_{n+1},a_n) - Q^{old}(s_n,a_n)\big]e_n(s,a)
		\end{multline}
		where
		\begin{equation}
			e_n(s,a)=\mathcal{I}_{ss_n}\mathcal{I}_{aa_n} + \begin{cases}
				\beta\lambda e_{n-1}(s,a), \hspace{0.9cm}a_n=\pi^*_{n-1}(s_n) \\
				0,\hspace{2.5cm} otherwise
			\end{cases}
		\end{equation}
		where $\mathcal{I}_{xy}$ is an identity indicator function which equals to 1 if $x=y$ and 0, otherwise. 
		
		In (23), the learning rate $(\beta)$ controls the Q value update rate. In addition, the discount factor  $(\gamma)$ decides how the immediate and future rewards are valued. $Q^{old}(s_n,a_n)$ is the old Q value, while $Q^{new}(s_n,a_n)$ is the new updated Q value. The term $\max\limits_{\!\!\underset{\scriptstyle a}{}}\ Q(s_{n+1},a_n)$ estimates the optimal future value, and $R_{n+1}$ is the reward received when transitioning from state $s_n$ to state $s_{n+1}$. In equation (24), the eligibility traces $e(s, a)$, controlled by the learning parameter $\lambda$ (similar to $\beta$), provide a higher update factor for recently revisited state-action pairs $(s, a)$, reinforcing their importance. The eligibility trace is cleared if the previous action $a_n$ is not greedy.

		In dynamic UAV networks with shifting topology and unstable links, fixed learning rates and discount factors prove insufficient. They should be adaptable and responsive to evolving network conditions. The suggested approach introduces adjustability to learning rate and discount factor. The dynamic learning rate is given as
		\begin{equation}
			\beta_{i}= \beta^{'}_{i}(\beta_{max}-\beta_{min})+\beta_{min}
		\end{equation}
			
			In (25), $\beta_{i}$ and $\beta^{'}_{i}$ are the dynamic and intermediate learning rates of $i^{th}$ UAV respectively. $\beta^{'}_{i}$ equals to $\big[\frac{1}{1+e^{-P^{cov}(i)}}\big]$, a function of coverage probability  $P^{cov}(i)$.  By defining $\beta_{max} = 1$ as the upper bound of the learning rate and $\beta_{min} = 0.01$ as the lower bound, we ensure initial rapid learning and continued exploration as the learning process advances. As  $P^{cov}$ increases, $\beta_{i}$ tends to decrease, allowing lower learning rates in well-covered, stable areas for steady learning. Conversely, regions with inadequate coverage and unstable links employ higher learning rates, facilitating rapid adaptation to changing network conditions.
			
			The number of candidate neighbour nodes determines the dynamic discount factor. If the number of candidate neighbour nodes is significant, indicating a well-connected network, the discount factor will approach 1, prioritizing future rewards during learning. Conversely, when the count of candidate neighbours is small, meaning a less connected network, the discount factor will approach 0, emphasizing immediate rewards. The  dynamic discount factor is defined as,
			\begin{equation}
				\gamma_i=\gamma^{'}_i(\gamma_{max}-\gamma_{min})+\gamma_{min}
			\end{equation}
		
			In (26),  $\gamma_i$ and $\gamma^{'}_i$ are the   dynamic and  intermediate discount factor of $i^{th}$ UAV respectively.  $\gamma^{'}_i$ is defined as $\frac{N_{c}(i)}{M}$,  the ratio of the current number of candidate neighbour nodes ($N_{c}(i)$), to the total count of nodes ($M$). A gradual decrease in the discount factor is achieved by setting $\gamma_{max} = 0.9$ and $\gamma_{min} = 0.1$ as the upper and lower bound, respectively. This choice balances immediate rewards and future considerations, enabling UAVs to adjust to evolving network conditions.

			\begin{figure}
				\centering
				\includegraphics[width=0.49\textwidth]{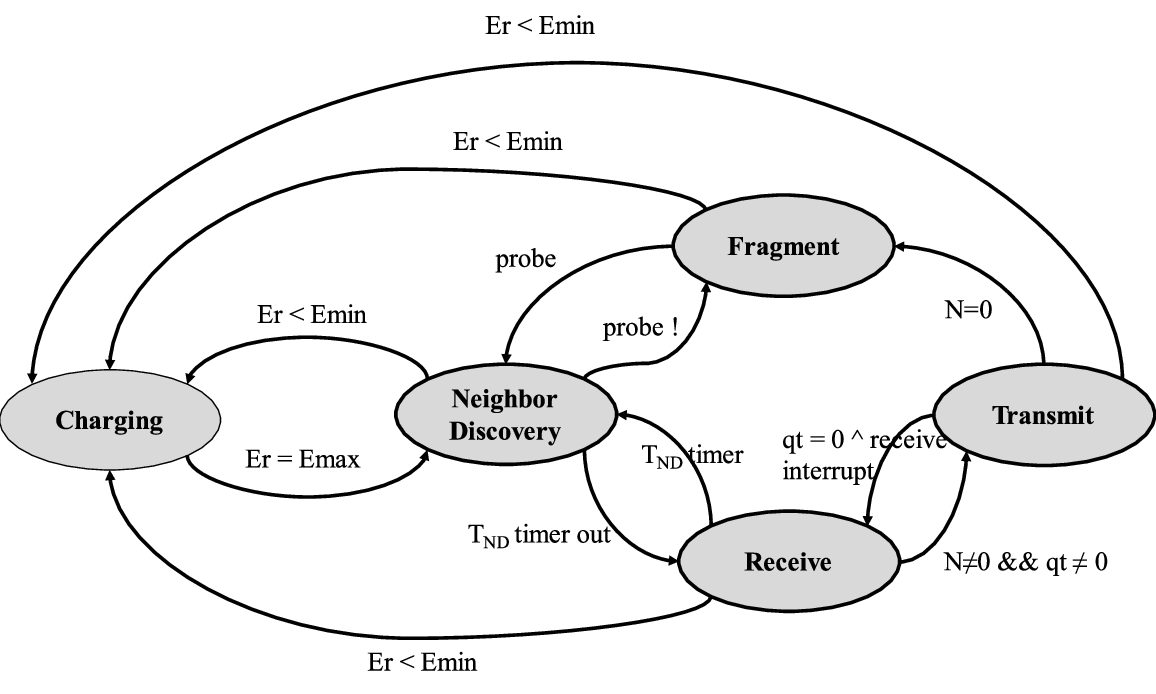}
				\caption{Operating modes of UAV.}
			\end{figure}
			
			The UAV’s operating modes include Neighbour Discovery, Receive, Transmit, and Charge. The mode transition diagram in Fig. 4 illustrates the sequence of  UAV modes. UAVs whose remaining energy levels exceed a specific threshold ($\mathcal{E}^{res} > \delta_{th}$) engage in Neighbor Discovery, Receive, and Transmit modes. If not, they transition to the Charge mode, relocating to designated ground charging points to replenish energy before returning to the network in Neighbour Discovery mode. In Neighbour Discovery, Receive, and Transmit modes, UAVs evaluate their residual energy levels and adjust the reward function in (21). Initially, UAVs commence with Neighbor Discovery until the \textit{HelloInterval} duration expires. Subsequently, they transition to the Receive mode, where received data is stored in the reception queue ($q_r$). Depending on the nature of the received data, whether it's surveillance data, \textit{ACKs}, or \textit{HelloMessage}, specific actions are performed, such as moving data to the transmission queue ($q_t$), adjusting the reward function in equation (21), or updating the network table. UAVs enter the Transmit mode only when there is data to relay ($q_t \neq 0$). In Transmit mode, UAVs evaluate $\mathcal{E}^{res}$ in (16), $\mathcal{P}^{rs}$ in (18), $\mathcal{P}^{cov}$ in (19) and $\mathcal{P}^{coll}$ in (20) and update the reward function to facilitate optimal decision-making for selecting the next-hop node using the proposed IQMR protocol. Neighbour discovery mode recommences either at the beginning of the \textit{HelloInterval} or when a possible next-hop node is unavailable due to network fragmentation ($N_H = 0$). 
		
			\textit{Algorithm 2} proposes  IQMR protocol for energy-efficient reliable data transfer via a multi-hop path. The IQMR protocol leverages information on residual energy, packet reception, coverage and collision probabilities and employs a learning method to update a Q-matrix. By selecting the forwarding node with the highest reward value and closest to the destination, the IQMR protocol maximizes successful data delivery in the network.
			
			\begin{algorithm}[htb!]
				\caption{: Improved Q-learning based Multi-hop Routing Protocol}\label{algo2}
				\textbf{Input:}Neighbour Table, Operating modes of UAV (neighbour discovery, receive, transmit, charge) \\
			
				\textbf{Output:} Energy efficient reliable multi-hop  data routing path for each UAV-GCS pair $(\pi^*)$\\
			
				\begin{algorithmic}[1]
					\STATE Establish and update the neighbour table (\textit {Algorithm 1})
					\WHILE{next hop node != GCS}
					\FOR{ each UAV $U$ in $U_1$ to $U_M$ }
					\STATE Compute distance ($r$) between neighbour UAVs
					\IF{$r\leq r_0$}
					\STATE $\mathcal{E}^{tx}(k,r)= \epsilon_{elec}k + \epsilon_{amp-fs}kr^2$ (13)
					\ELSE
					\STATE $\mathcal{E}^{tx}(k,r)=\epsilon_{elec}k +\epsilon_{amp-mp}kr^4$ (14)
					\ENDIF 
					\STATE Compute the residual energy $\mathcal{E}^{res}$ (16)
					\IF{$\mathcal{E}^{res} < \delta_{th}$ }
					\STATE Mode = \textit{Charge}
					\ELSE
					\IF{Mode is \textit{Neighbour Discovery}}
					\STATE Update the neighbour table  (\textit{Algorithm 1})
					\ELSIF{Mode is \textit{Receive}}
					\IF{Data is \textit{HelloMessage}}
					\STATE Update the neighbour table (\textit{Algorithm 1})
					\ELSIF{Data is acknowledgement (\textit{ACKs})}
					\STATE Update packet reception status ($\mathcal{P}^{rs}$) (18)
					\ELSIF {Data is surveillance information}
					\STATE Update transmission queue ($q_t$)
					\ENDIF
					\IF {reception queue is empty ($q_r=0$) \& transmission queue is not empty($q_t \neq 0 $)}
					\STATE Mode is \textit{Transmit}
					\ELSE
					\STATE Mode is \textit{Receive}
					\ENDIF
					\ELSIF {Mode is \textit{Transmit}}
					\STATE Compute packet reception status ($\mathcal{P}^{rs}$) (18)
					\STATE Compute coverage probability ($\mathcal{P}^{cov}$) (19)
					\STATE Compute collision probability ($\mathcal{P}^{coll}$) (20) 
					\STATE Measure reward ($R$) (21)
					\STATE Update Q-matrix.
					\STATE Return  optimal next hop node ($\pi^*$)
					\IF{transmission queue is not empty ($q_r \neq 0$) or reception queue is  empty ($q_t= 0 $)}
					\STATE Mode is \textit{Receive}
					\ELSE
					\STATE Mode is \textit{Transmit}
					\ENDIF
					\ENDIF
					\ENDIF
					\ENDFOR
					\ENDWHILE
				\end{algorithmic}
			\end{algorithm}
			\section{Results and Discussion}
			This section analyses the performance of the proposed IQMR protocol through extensive simulation. First, we illustrate the IQMR protocol decision process. Second, we discuss the behavior of the proposed protocol for varying system parameters (residual energy, packet reception status, coverage, and collision probabilities). Third, we compare the IQMR protocol with the existing routing algorithms, namely Q-FANET\cite{QFANET}, QMR \cite{QMR} and QTAR \cite{QTAR} for (i) energy efficiency (rate of decrease in residual energy) and (ii) reliability (number of data packets received at  GCS).
			\subsection{Simulation Environment}
			UAVs are randomly distributed in a 3D  space, spanning a  radius of 1000 m and a maximum height of 300 m. The UAVs' altitudes vary between 100-300 m, with speeds ranging from 10-30 m/s. Each UAV has a uniform transmission power of 1 W and a radio range of 250 m. In each iteration, the UAVs can transmit directly to GCS or nearby UAVs within the radio range. Each UAV device has a capacity of 207792 (11.1 V, 5200 mAh \cite{Gupta}) Joules of energy. Table I summarizes the detailed parameters considered for simulation. 
			\begin{table}[h]
				\begin{center}
					\caption{List of simulation parameters}
					\begin{tabular}{@{}ll@{}}
						\hline
						\vspace{1pt}
						\textbf {Parameter} & \textbf{Specification}\\ 
						\hline
						network area (cylindrical) & R=1000 m, H=300 m \\
						number of UAVs (${M}$) & 50\\
						number of GCS  & 1\\
						UAV operating altitude $[h_{max}, h_{min}]$ & 100-300 m\\
						UAV transmission radius & 250 m \\
						UAV transmit power & 1 W\\
						UAV velocity $[v_{max}, v_{min}]$ & 10-20 m/s \\
						UAV movement direction $[d_{max}, d_{min}]$ & $[-\pi/2, \pi/2]$\\
						mobility model & 3D Gauss Markov Mobility  Model\\
						pause time $[t_{min}, t_{max}]$ & 1-20 s\\
						channel propagation model & Nakagami-$m$ small scale fading\\
						SIR threshold ($\theta$)  & 0 dB \\
						$\mathcal{E}^{full}$ \cite{Gupta} & 207792 J(11.1V, 5200 mAh)\\
						$\delta_{th}$ & 1000 J\\
						$r_0$& 100 m\\
						$\mathcal{E}_{elec}$ \cite{eng-eqs} & 50 nJ\\
						$\mathcal{E}_{amp-fs}$ \cite{eng-eqs}&41 $\mu$J\\
						$\mathcal{E}_{amp-mp}$ \cite{eng-eqs}& 100 pJ\\
						$\mathcal{E}_{payload}$ \cite{eng-eqs}& 0.217 kW/kg\\
						$w$ \cite{engfly}& 5 kg \\
						$\mathcal{E}_{hover}$ \cite{engfly}& 0.185 W\\
						$\mathcal{E}_{density}$ \cite{engfly}& 650 kJ/kg\\
						$\xi_x$ = $\xi_y$ & 3m\\
						traffic type& CBR\\
						CBR rate & 2Mbps\\
						packet size & 150 bytes\\
						network simulator&MATLAB\\
						\hline
					\end{tabular}
				\end{center}
			\end{table}
			
			\subsection{Performance of IQMR routing decision}
			Fig. 5 shows a simple UAV network topology consisting of a source node (UAV 1), a destination node (GCS), and eight relay nodes (UAVs 2  to 9). Let at time $n$ the source node have a data packet to forward. The source node formulates its neighbour set following the exchange of \textit{HelloMessage}. The candidate neighbour set of UAV 1 includes UAVs 2,3,4 and 9 (as discussed in \textit{Algorithm 1}).
		
			\begin{figure}[]
				\centering
				\subfigure[]{\includegraphics[width=0.49\linewidth]{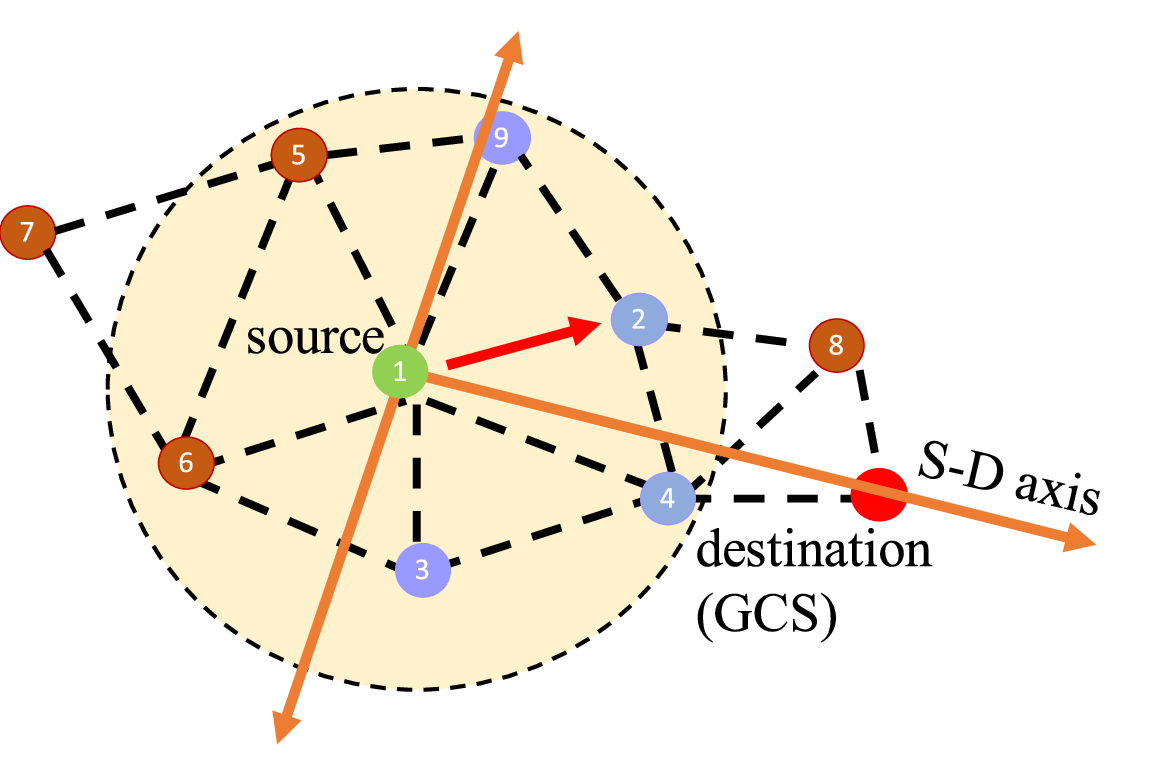}} 
				\subfigure[]{\includegraphics[width=0.49\linewidth]{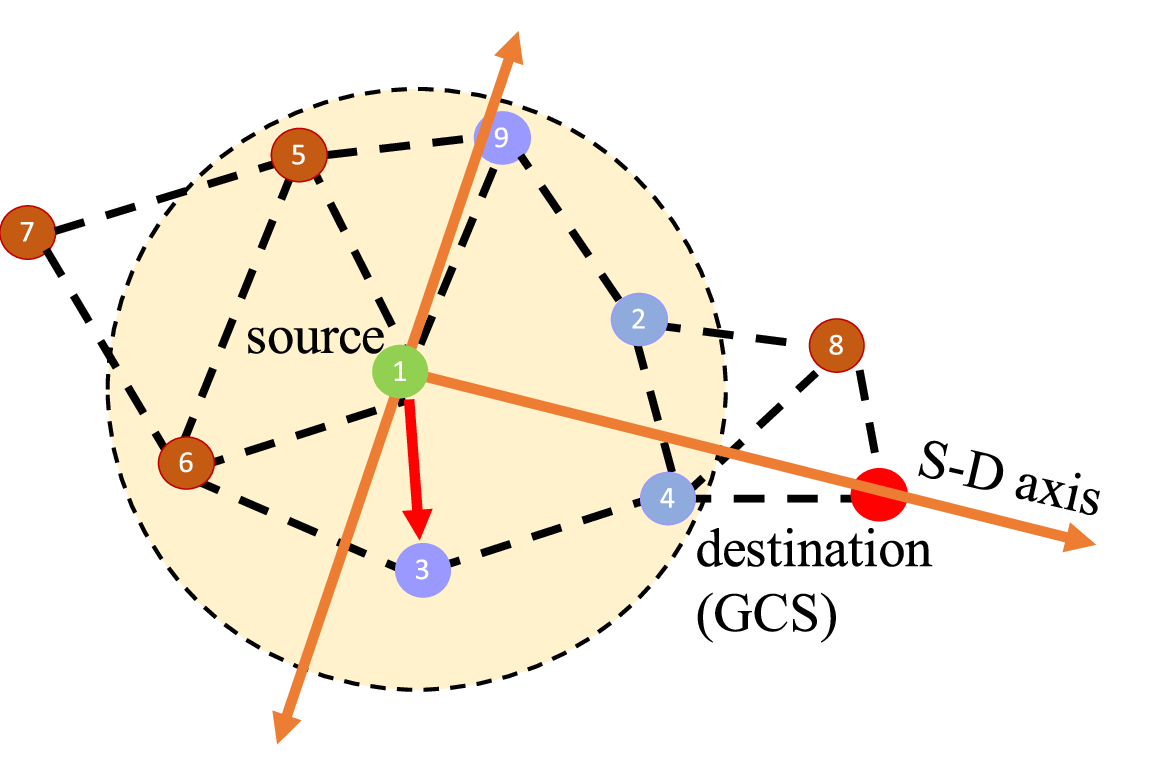}} 
				\caption{Illustration of next-hop node selection for a static network  with one source (UAV 1) and eight relay nodes (UAVs 2 to 9) and destination node (GCS); S-D axis indicating source-destination node axis. Representation of nodes are as green-source node, blue-candidate neighbour nodes, brown- neighbour and non neighbour nodes, red-destination node, grey-candidate nodes that do not satisfy routing constraints.}
				%
			\end{figure} 
		
			At the time $n$, the Q-values corresponding to the candidate neighbour set [UAV 1, UAV 3, UAV 4, UAV 9] are [0.90, 0.65, 0.85,0.65]. The source node, in general, forwards the packet to the node with the highest Q-value, i.e., UAV 2 (Fig.5a). However, when two candidate neighbours have the same Q-value, then the one less divergent from the source-destination node (S-D) axis is selected as the next forwarder. Assume node 2 and node 4  do not satisfy the constraints (9),(10), or  (11), then the source node does not consider the UAVs 2 and 4 for selection as next-hop nodes. Instead, the source node routes the data to node 3 (Fig.5b). Though the Q-values of nodes 9 and 3 are equal, the latter is selected as the next-hop node since it is comparatively less divergent from the S-D  axis.
			\subsection{Impact of System Parameters on IQMR protocol}
			This section discusses the impact of  varying system parameters such as residual energy, packet reception status, coverage and collision probabilities of the proposed IQMR protocol. We then study the system performance when the UAVs are fragmented from the network. In what follows, in Figs. 6 to 13, the $x$-axis represents the number of episodes of data transmission, and the $y$-axis represents the cumulative reward.
			\subsubsection{Residual Energy}
			The residual energy of the network is the sum of the remaining energies of all the UAVs after each episode of data transmission. Initially, the UAVs are powered with 11.1 V and 5200 mAh battery. The energy expended for data dissemination follows the distance-dependent energy consumption (described in Section III). This section analyzes the variation of residual energy over the simulated UAV network. Figs. 6 and 7 illustrate the system's cumulative reward behavior when the residual energy of a few UAVs falls below a certain threshold (100 J) simultaneously. Figs. 6a-6d depicts the use case of $20\%$  UAVs (randomly chosen) within the coverage range of the transmitting UAV  with residual energy less than 100 J. When UAVs run out of charge, their ability to complete the surveillance and relay the data are compromised, leading to unstable rewards obtained by the system (indicated as non-uniform region in the cumulative reward graph). The plateauing in the cumulative reward graph represents the recovery phase, where the system is relatively stable to disseminate information. The system performance is observed over varying learning rates 1 (Fig. 6a), 0.5 (Fig. 6b), 0.1 (Fig. 6c) and 0.01 (Fig. 6d). The convergence rate of the proposed IQMR protocol to a stable policy depends on the learning rate, which determines how the algorithm is updated. A higher learning rate implies larger updates, while a lower learning rate leads to smaller updates. Hence as depicted in the graph, the system with a higher learning rate settles down faster and vice versa.

			\begin{figure}[]
				\centering
				\subfigure[]{\includegraphics[width=0.24\textwidth]{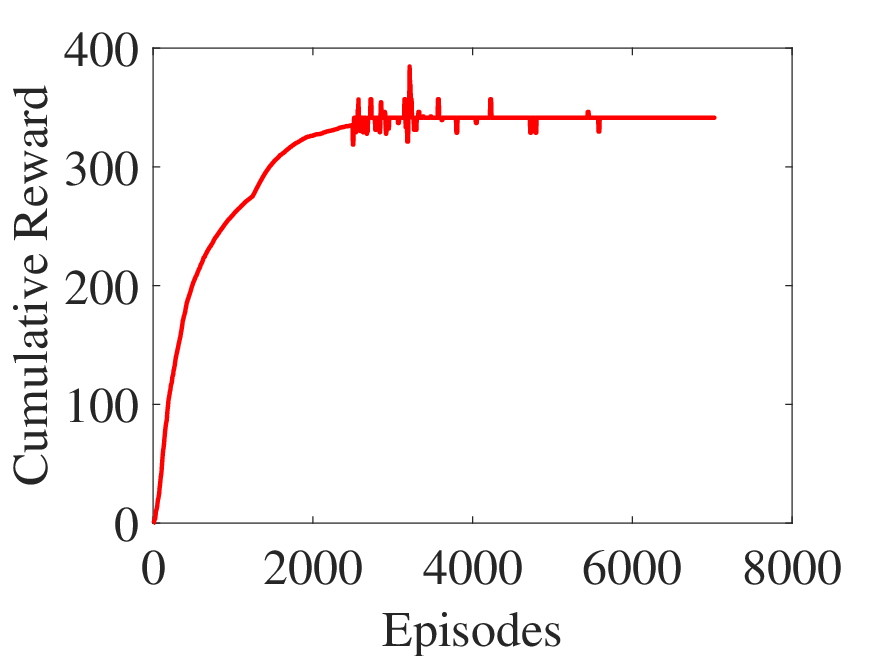}} 
				\subfigure[]{\includegraphics[width=0.24\textwidth]{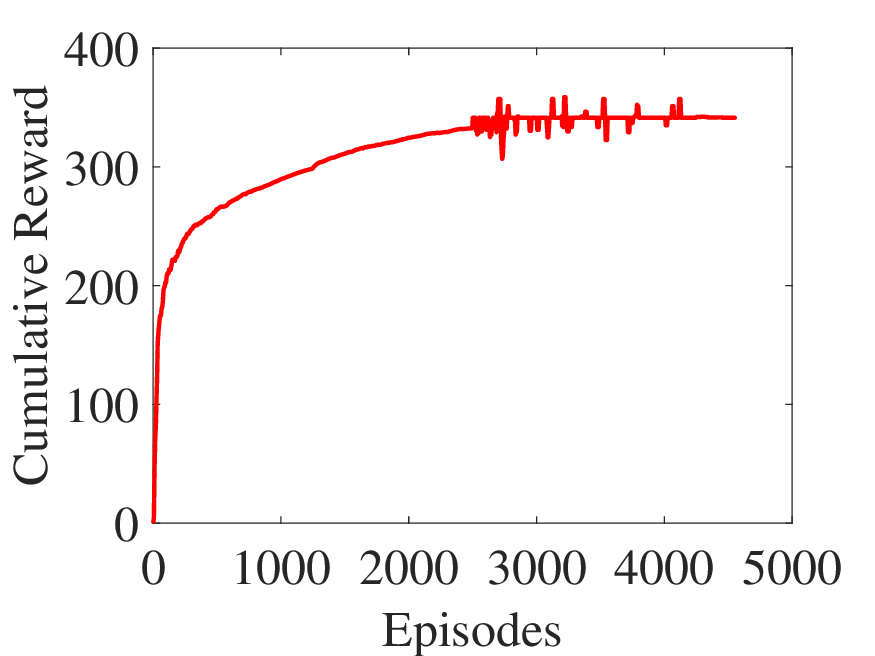}} 
				\subfigure[]{\includegraphics[width=0.24\textwidth]{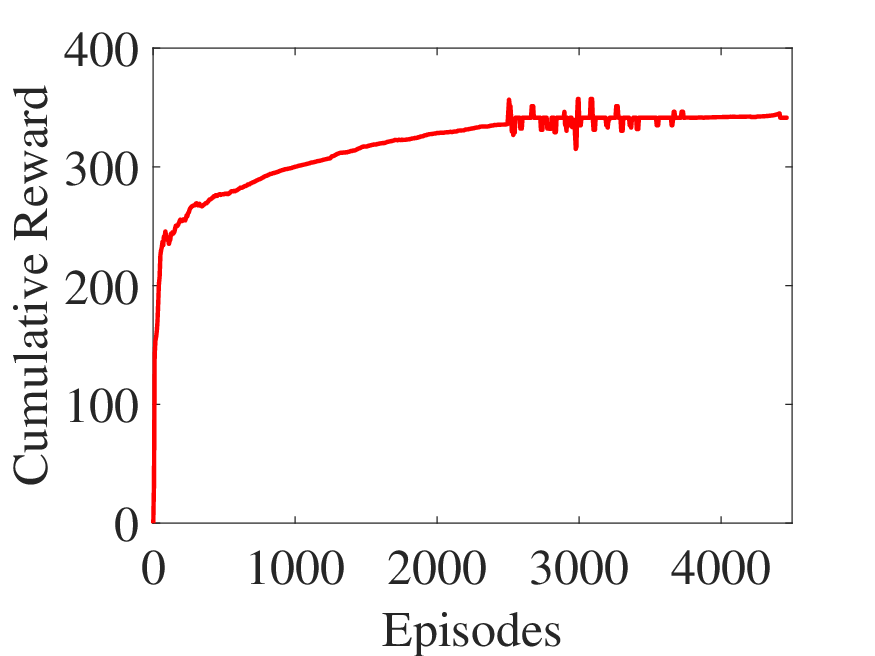}}
				\subfigure[]{\includegraphics[width=0.24\textwidth]{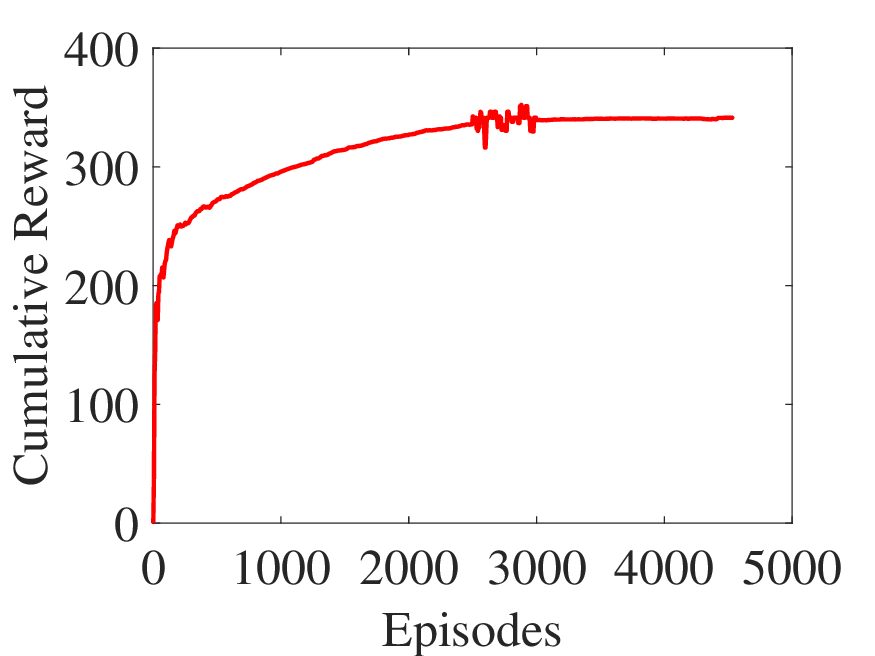}}
				\caption{Impact on cumulative reward  as a function of residual energy for varying learning rates (a) 0.01 (b) 0.1 (c) 0.5 (d) 1.}
				%
			\end{figure} 
			Fig. 7 details the system behavior as a function of residual energy and Q-value. The UAVs are arranged in descending order based on their Q-values and then divided into two sets. Set 1 comprises the top half of UAVs with higher Q-values, while Set 2 includes the remaining half with lower Q-values. This division allows for evaluating the impact of different subsets of UAVs on system performance. The Q-value indicates the suitability of a UAV for forwarding data packets. UAVs with higher Q-values are considered more reliable for routing decisions. UAVs with higher Q-values exiting the network due to insufficient operational energy ($ \mathcal{E}^{res}< 100$ J) substantially influence the cumulative reward. Their departure disrupts the optimal data flow within the network, rendering the absence of UAVs with higher Q-values more impact on the cumulative reward (Fig. 7a) compared to UAVs with lower Q-values (Fig. 7b). 
			
			\begin{figure}[]
				\centering
				\subfigure[]{\includegraphics[width=0.24\textwidth]{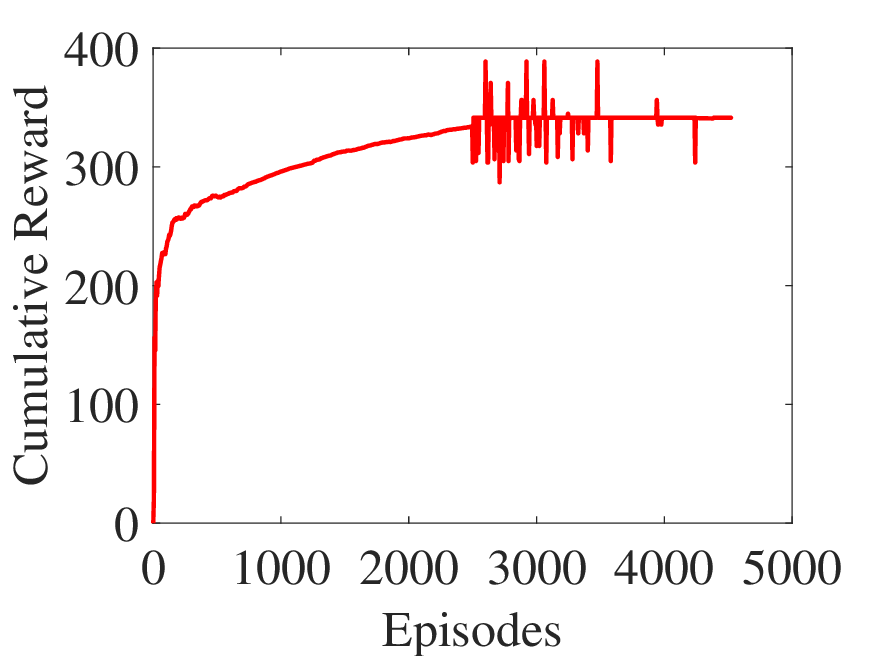}} 
				\subfigure[]{\includegraphics[width=0.24\textwidth]{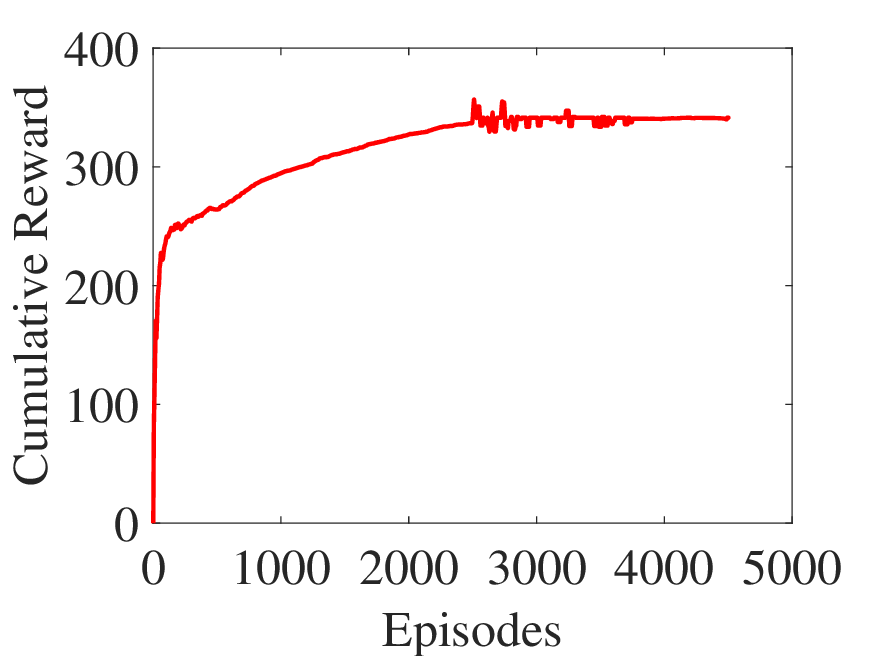}} 
				\caption{Variation of  cumulative reward  as a function of residual energy and Q-values.}
			\end{figure} 
			\subsubsection{Packet Reception Status}
			Fig. 8a details the influence of layers 2 and 3 \textit{ACKs} (\textit{L2} and \textit{L3 ACKs}) on the cumulative reward. \textit{L2 ACKs} are important for next-hop communication while \textit{L3 ACKs} have a broader implication for end-to-end delivery, routing decision and  network-wide operation. Therefore higher cumulative reward is achieved with the reception of \textit{L3 ACKs} than \textit{L2 ACKs}. 
			\subsubsection{Coverage Probability}
				Fig. 8b describes the impact of the SIR threshold ($\theta$) on the cumulative reward. The parameter $\theta$ is crucial in determining the $\mathcal{P}^{cov}$. A higher $\theta$ value leads to a decrease in $\mathcal{P}^{cov}$, while a lower $\theta$ value increases $\mathcal{P}^{cov}$. When $\theta$ is set to a higher value, the receiver requires a stronger signal than the interference level for successful communication. As a result, $\mathcal{P}^{cov}$ decreases because only a few UAVs close to the transmitter or experiencing low interference levels can meet the stringent SIR requirement. On the other hand, when $\theta$ is set to a lower value, the receiver can tolerate a weaker signal relative to the interference level, leading to higher $\mathcal{P}^{cov}$. In this case, UAVs located farther away or experiencing higher interference levels have a greater chance of meeting the less stringent SIR requirement. Thus, increasing the value of $\theta$ limits the pool of potential next-hop nodes for data transmission, leading to a decline in the cumulative reward, as illustrated in Fig.9
				\subsubsection{Collision Probability}
				Fig. 8c represents the impact of inter-UAV distance ($r$) on the behavior of the cumulative reward of the proposed system. The $\mathcal{P}^{col}$  can be effectively minimized by maintaining adequate separation between UAVs, thus, enhancing reliable information transmission. Therefore as $r$ increases, the  $\mathcal{P}^{col}$ decreases, positively influencing the cumulative reward.

				\begin{figure}[]
					\centering
					\subfigure[]{\includegraphics[width=0.24\textwidth]{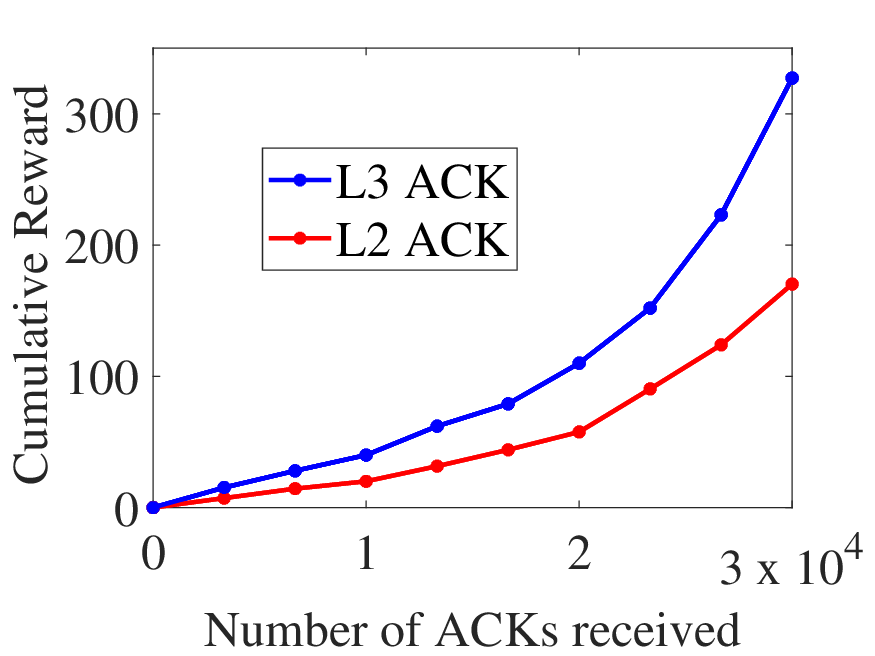}} 
					\subfigure[]{\includegraphics[width=0.24\textwidth]{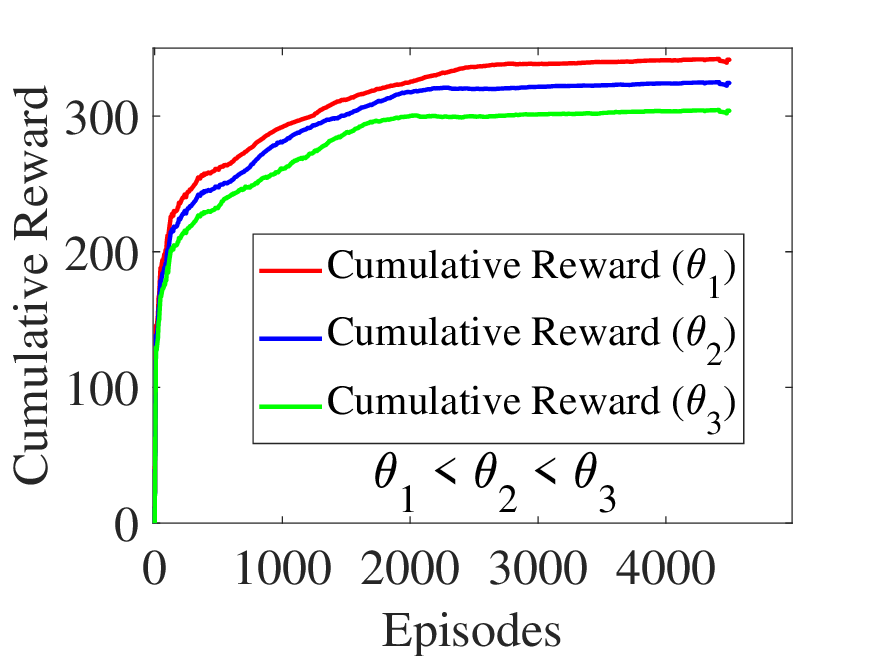}} 
					\subfigure[]{\includegraphics[width=0.24\textwidth]{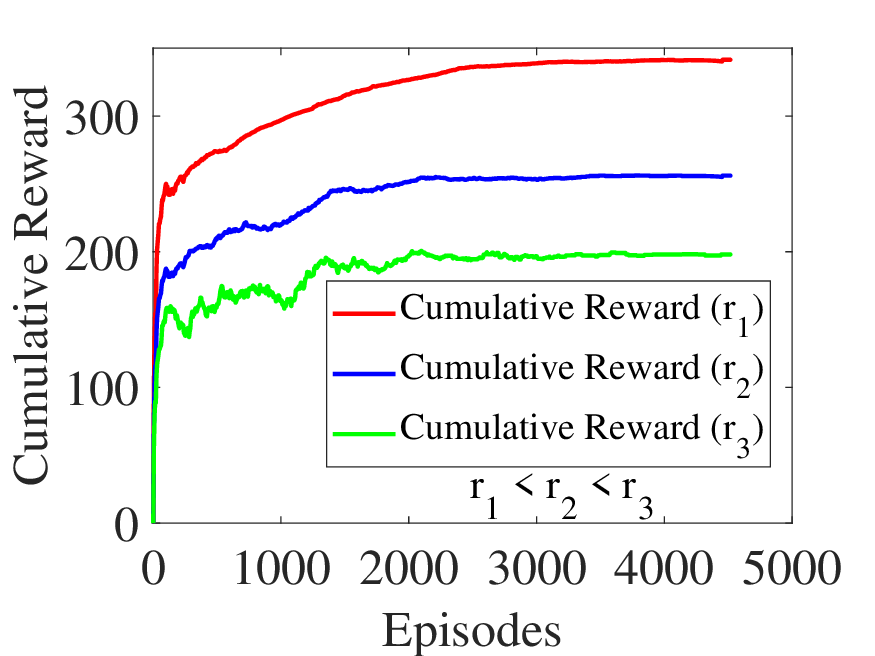}}
					\caption{Impact on cumulative reward for variation in (a) \textit{L2} and \textit{L3} ACKs  (b) SIR (c) inter-UAV distance (\textit{r}) }
					%
				\end{figure} 
			
					\subsubsection{Network Fragmentation}
					This section analyzes the effect of node fragmentation on the cumulative reward. A proportion of the UAVs are assumed to be fragmented simultaneously and remain out of the network for a fixed duration. The nodes then join the network back together or in batches. Figs. 9 and 10 show cumulative reward variations when higher Q-valued UAVs within the coverage range of transmitting UAV fragments from the network for 400ms and 200ms respectively, with all (Figs. 9a, 10a) or a quarter (Figs. 9b, 10b) of them rejoining over the next 10ms. Fig. 11 enumerates the impact on the cumulative reward when $20\%$  (randomly chosen) of UAVs within the coverage range of transmitting UAV fragments for 200ms, with all (Fig. 11a) or a quarter (Fig. 11b) of them rejoining over the next 10ms. Fragmentation of higher Q-valued UAVs has significant implications on routing performance, such as decreased routing efficiency, network resilience, and increased communication overhead. Conversely,  simultaneous rejoining (Figs. 9a, 10a, 11a) causes congestion, collisions, and sub optimal routing. In contrast, gradual rejoining (Figs. 9b, 10b, 11b) facilitates smoother assimilation and faster system stabilization.

					\begin{figure}[ht]
						\centering
						\subfigure[]{\includegraphics[width=0.24\textwidth]{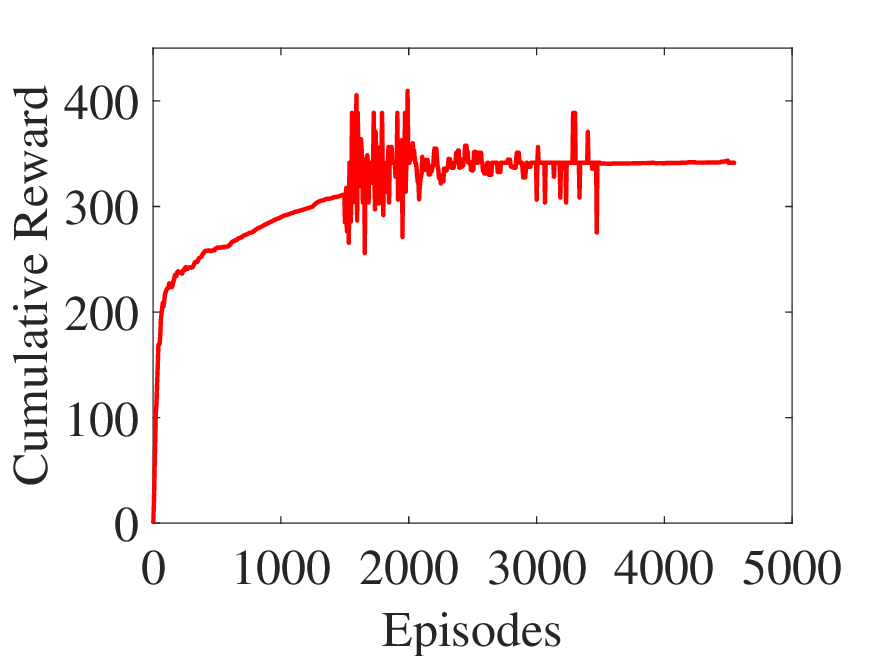}} 
						\subfigure[]{\includegraphics[width=0.24\textwidth]{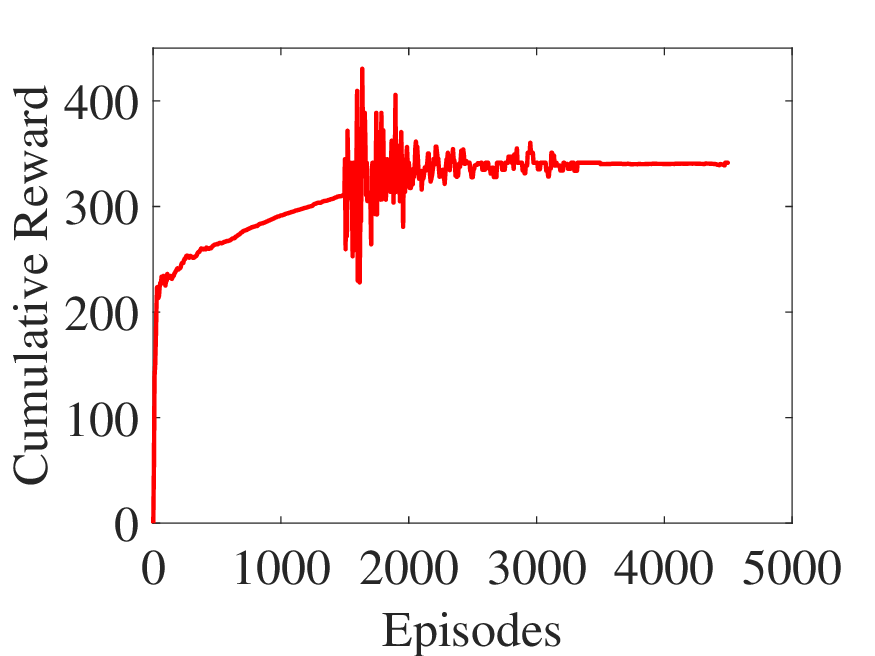}} 
						\caption{Impact on the cumulative reward when higher Q-valued UAVs  are partitioned from the network for 400 ms, and  (a) all (b)  quarter join the network back over the next 10 ms.}
					\end{figure} 
					\begin{figure}[]
						\centering
						\subfigure[]{\includegraphics[width=0.24\textwidth]{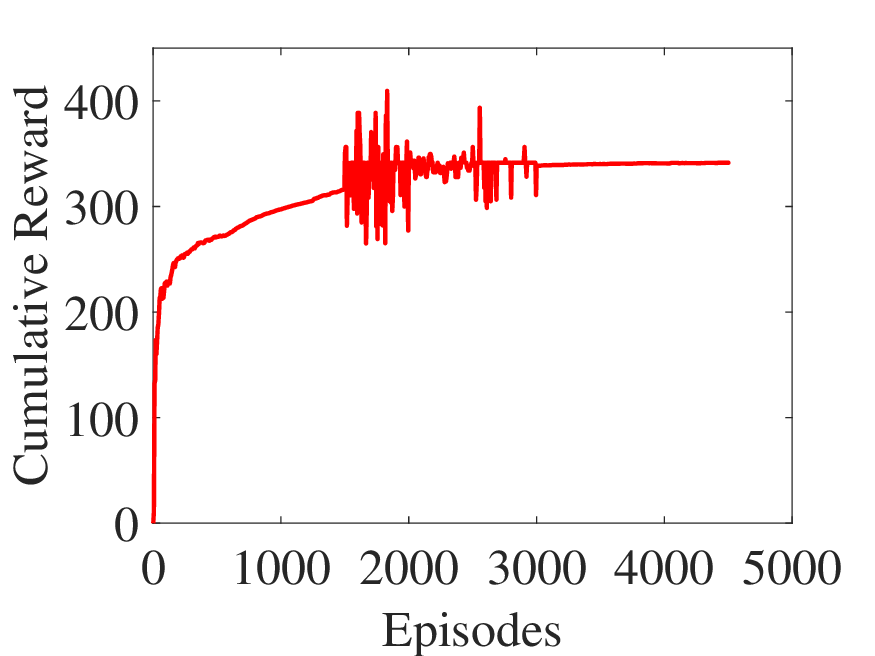}} 
						\subfigure[]{\includegraphics[width=0.24\textwidth]{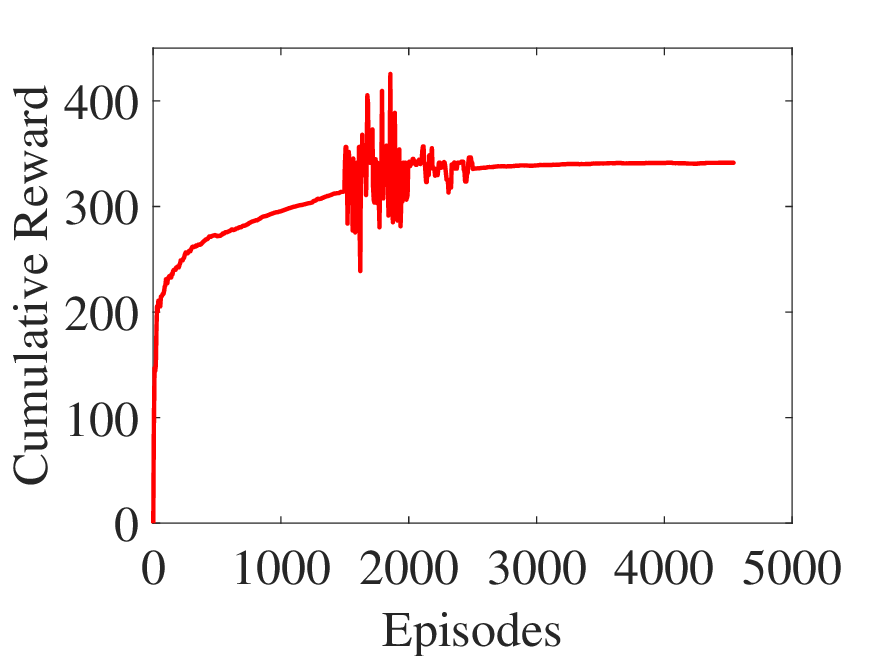}} 
						\caption{Impact on the cumulative reward when higher Q-valued UAVs  are partitioned from the network for 200 ms, and  (a) all (b)  quarter join the network back over the next 10 ms. }
					\end{figure} 
					\begin{figure}[]
						\centering
						\subfigure[]{\includegraphics[width=0.24\textwidth]{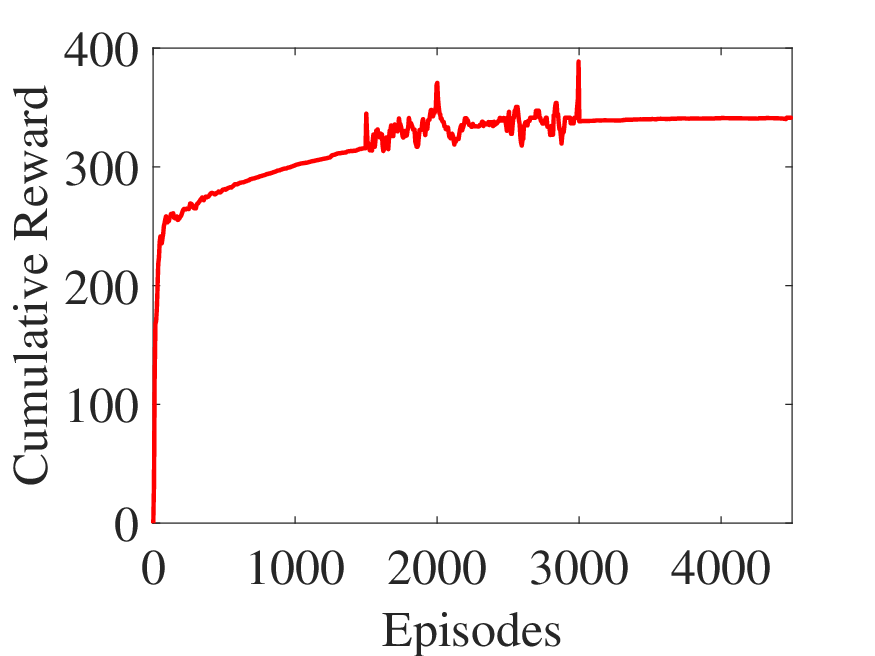}} 
						\subfigure[]{\includegraphics[width=0.24\textwidth]{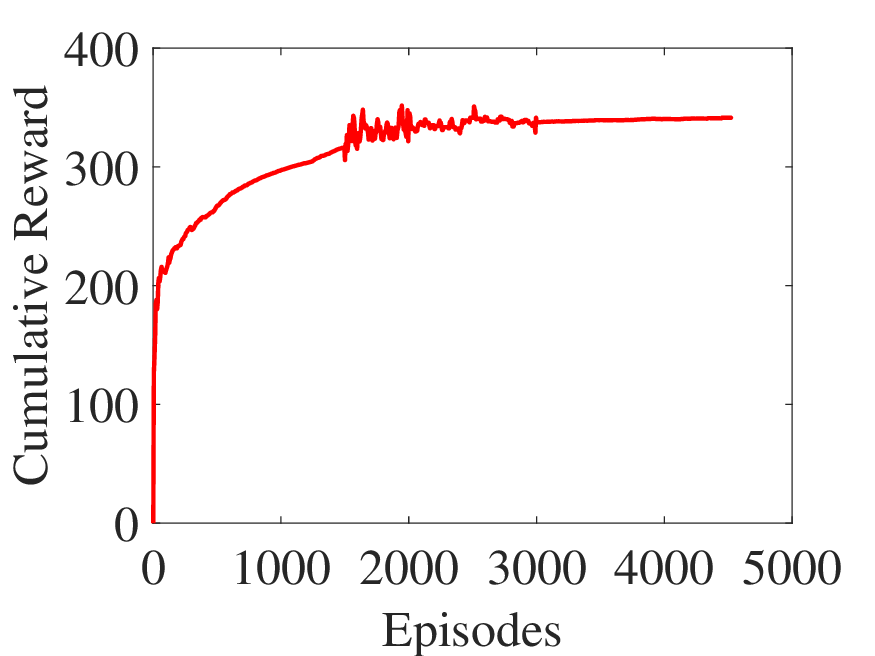}} 
						\caption{Impact on the cumulative reward when $20\%$ UAVs (randomly chosen) are partitioned from the network for 200 ms, and  (a) all (b) quarter join the network back over the next 10 ms.}
					\end{figure}

					\subsection{ Performance Comparison: IQMR with Existing Protocols}
					This section compares and analyses the performance of the proposed IQMR protocol with the existing routing methods Q-FANET, QMR and QTAR. 
					\subsubsection{IQMR Energy Efficiency Analysis}
					As depicted in Fig. 12a, the IQMR protocol demonstrates superior energy consumption compared to Q-FANET, QMR, and QTAR. For the QTAR and Q-FANET protocols, the residual energy reduces to 0 after 2000 and 5000 episodes, respectively. In contrast, by the 8000 episode, QMR retains only 4.07\% of its initial energy, whereas IQMR retains 36.34\%. Thus IQMR attains 32.27\%, 36.35\%, 36.35\% higher energy consumption efficiency than QMR, QFANET and QTAR. IQMR's adeptness in selecting energy-efficient UAVs as next-hop nodes ensures balanced energy consumption across the network. Furthermore, IQMR's propensity for selecting higher-quality links translates to reduced re-transmissions and elevated energy efficiency
					\begin{figure}[]
						\centering
						\subfigure[]{\includegraphics[width=0.24\textwidth]{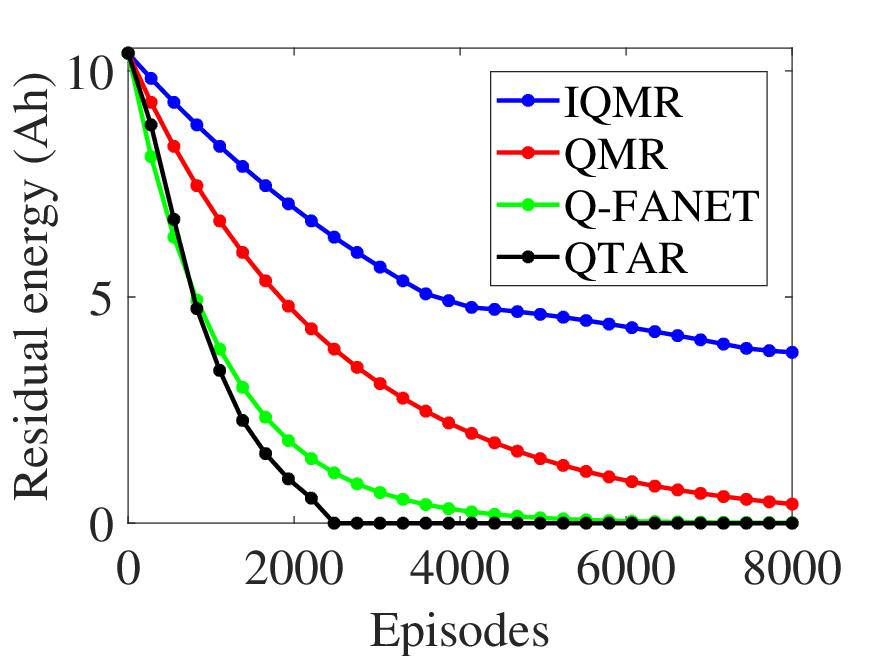}} 
						\subfigure[]{\includegraphics[width=0.24\textwidth]{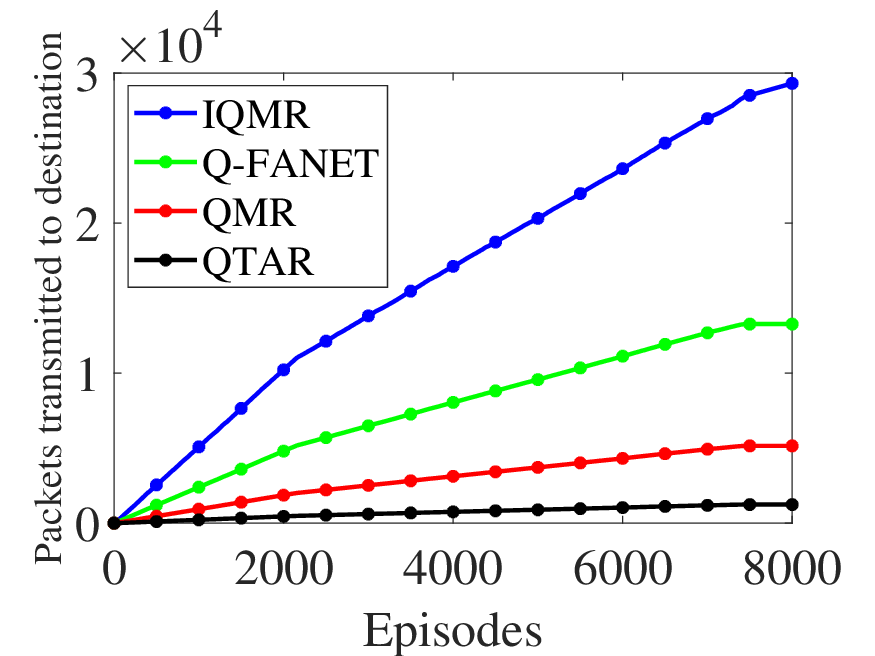}} 
						\caption{Illustration of (a) residual energy and (b) data throughput over varying episodes of data transmission.}
					\end{figure} 
					\subsubsection{IQMR Data Throughput Analysis}
					Fig. 12b illustrates the data throughput quantifying the number of successfully transmitted packets to the GCS. The total number of transmitted packets in the network is  $3\times10^4$. At the 8000-episode mark, IQMR successfully transmits about $2.9\times10^4$ packets to the GCS. While QMR, Q-FANET,  and QTAR manage approximately $1.3\times10^4$, $0.5\times10^4$ and $.12\times10^4$ successful data packet transfers to the GCS. Thus, IQMR achieves 53.3\%, 80.35\% and 93.36\%  higher data throughput than QMR, Q-FANET, and QTAR, respectively. IQMR achieves substantial data throughput improvement by selecting optimal links, reducing data loss through collision avoidance, and successful reception assessment.

					\section{Conclusion}
			
					This study presents a novel IQMR protocol for multi-hop data routing in dynamic UAV networks. The protocol involves neighbor discovery, network parameter estimation for energy-efficient and reliable communication, and adaptive routing using improved Q-learning. IQMR ensures energy efficiency, connectivity and inter-UAV collision prevention, and network restoration post-UAV fragmentation without prior UAV trajectory planning. The study analyzes the behavior of cumulative reward to changing system parameters, such as residual energy, coverage, collision probabilities, packet reception status, and fragmentation. A comparison of IQMR with QMR, Q-FANET, and QTAR protocols demonstrates IQMR's superior energy efficiency and data throughput. IQMR achieves energy consumption efficiency improvements of 32.27\%, 36.35\%, and 36.35\% over QMR, Q-FANET, and QTAR, along with significantly higher data throughput enhancements of 53.3\%, 80.35\%, and 93.36\% over Q-FANET, QMR, and QTAR.
					
					Future extensions to the IQMR protocol include integrating mobility prediction, addressing scalability and large-scale deployment, and integrating edge computing capabilities.

				
					\bibliographystyle{IEEEtran}
					\bibliography{References}

\begin{thebibliography}{10}
\providecommand{\url}[1]{#1}
\csname url@samestyle\endcsname
\providecommand{\newblock}{\relax}
\providecommand{\bibinfo}[2]{#2}
\providecommand{\BIBentrySTDinterwordspacing}{\spaceskip=0pt\relax}
\providecommand{\BIBentryALTinterwordstretchfactor}{4}
\providecommand{\BIBentryALTinterwordspacing}{\spaceskip=\fontdimen2\font plus
\BIBentryALTinterwordstretchfactor\fontdimen3\font minus
  \fontdimen4\font\relax}
\providecommand{\BIBforeignlanguage}[2]{{%
\expandafter\ifx\csname l@#1\endcsname\relax
\typeout{** WARNING: IEEEtran.bst: No hyphenation pattern has been}%
\typeout{** loaded for the language `#1'. Using the pattern for}%
\typeout{** the default language instead.}%
\else
\language=\csname l@#1\endcsname
\fi
#2}}
\providecommand{\BIBdecl}{\relax}
\BIBdecl

\bibitem{Mozaffari}
M.~Mozaffari, W.~Saad, M.~Bennis, Y.-H. Nam, and M.~Debbah, ``A tutorial on
  uavs for wireless networks: Applications, challenges, and open problems,''
  \emph{IEEE Communications Surveys \& Tutorials}, vol.~21, no.~3, pp.
  2334--2360, 2019.

\bibitem{HelpSky}
M.~Erdelj, E.~Natalizio, K.~R. Chowdhury, and I.~F. Akyildiz, ``Help from the
  sky: Leveraging uavs for disaster management,'' \emph{IEEE Pervasive
  Computing}, vol.~16, no.~1, pp. 24--32, 2017.

\bibitem{surveillance}
V.~A. Memos and K.~E. Psannis, ``Uav-based smart surveillance system over a
  wireless sensor network,'' \emph{IEEE Communications Standards Magazine},
  vol.~5, no.~4, pp. 68--73, 2021.

\bibitem{paradigm}
B.~Alzahrani, O.~S. Oubbati, A.~Barnawi, M.~Atiquzzaman, and D.~Alghazzawi,
  ``Uav assistance paradigm: State-of-the-art in applications and challenges,''
  \emph{Journal of Network and Computer Applications}, vol. 166, p. 102706,
  2020.

\bibitem{UAVdelivery}
M.~Sajid, H.~Mittal, S.~Pare, and M.~Prasad, ``Routing and scheduling
  optimization for uav assisted delivery system: A hybrid approach,''
  \emph{Applied Soft Computing}, vol. 126, p. 109225, 2022.

\bibitem{relaying}
M.~Nauman~Bashir, S.~Iqbal, and K.~Mohamad~Yusof, ``Design principles for
  cooperative relaying on uavs-based fanet,'' in \emph{2022 Advances in Science
  and Engineering Technology International Conferences (ASET)}, 2022, pp. 1--6.

\bibitem{Gupta}
L.~Gupta, R.~Jain, and G.~Vaszkun, ``Survey of important issues in uav
  communication networks,'' \emph{IEEE Communications Surveys \& Tutorials},
  vol.~18, no.~2, pp. 1123--1152, 2016.

\bibitem{mavanet}
O.~S. Oubbati, M.~Atiquzzaman, P.~Lorenz, M.~H. Tareque, and M.~S. Hossain,
  ``Routing in flying ad hoc networks: Survey, constraints, and future
  challenge perspectives,'' \emph{IEEE Access}, vol.~7, pp. 81\,057--81\,105,
  2019.

\bibitem{disaster-area}
S.~A. Hussain, N.~Ahmad, L.~A. Latiff, S.~B. Mohammed~Ahmed, and N.~Mohamed,
  ``Disaster area network expansion using drones based ad-hoc cellular
  communications,'' in \emph{2021 IEEE Symposium On Future Telecommunication
  Technologies (SOFTT)}, 2021, pp. 22--27.

\bibitem{publicsafety}
K.~Ali, H.~X. Nguyen, Q.-T. Vien, P.~Shah, and M.~Raza, ``Deployment of
  drone-based small cells for public safety communication system,'' \emph{IEEE
  Systems Journal}, vol.~14, no.~2, pp. 2882--2891, 2020.

\bibitem{LAROD}
M.~Bilal, Z.~Hassan, M.~Ayub, I.~Ahmad, and R.~Odarchenko, ``Classification of
  routing protocols for fants,'' in \emph{2019 IEEE International
  Scientific-Practical Conference Problems of Infocommunications, Science and
  Technology}, 2019, pp. 350--358.

\bibitem{LCAD}
C.-M. Cheng, P.-H. Hsiao, H.~T. Kung, and D.~Vlah, ``Maximizing throughput of
  uav-relaying networks with the load-carry-and-deliver paradigm,'' in
  \emph{2007 IEEE Wireless Communications and Networking Conference}, 2007, pp.
  4417--4424.

\bibitem{OLSR}
M.~H. Tareque, M.~S. Hossain, and M.~Atiquzzaman, ``On the routing in flying ad
  hoc networks,'' in \emph{2015 Federated Conference on Computer Science and
  Information Systems (FedCSIS)}, 2015, pp. 1--9.

\bibitem{LADTR}
M.~Y. Arafat and S.~Moh, ``Location-aided delay tolerant routing protocol in
  uav networks for post-disaster operation,'' \emph{IEEE Access}, vol.~6, pp.
  59\,891--59\,906, 2018.

\bibitem{QMR}
J.~Liu, Q.~Wang, C.~He, K.~Jaffrès-Runser, Y.~Xu, Z.~Li, and Y.~Xu,
  ``Qmr:q-learning based multi-objective optimization routing protocol for
  flying ad hoc networks,'' \emph{Computer Communications}, vol. 150, pp.
  304--316, 2020.

\bibitem{QFANET}
L.~A.~L. {da Costa}, R.~Kunst, and E.~{Pignaton de Freitas}, ``Q-fanet:improved
  q-learning based routing protocol for fanets,'' \emph{Computer Networks},
  vol. 198, p. 108379, 2021.

\bibitem{QTAR}
M.~Y. Arafat and S.~Moh, ``A q-learning-based topology-aware routing protocol
  for flying ad hoc networks,'' \emph{IEEE Internet of Things Journal}, vol.~9,
  no.~3, pp. 1985--2000, 2022.

\bibitem{DCR}
M.~Erdelj, M.~Król, and E.~Natalizio, ``Wireless sensor networks and multi-uav
  systems for natural disaster management,'' \emph{Computer Networks}, vol.
  124, pp. 72--86, 2017.

\bibitem{BATMAN}
N.~N. D.~Johnson and C.~Aichel, ``Simple pragmatic approach to mesh routing
  using batman,'' in \emph{2nd IFIP Int. Symp. Wireless Commun. Inf. Technol.
  Developing Countries}, 2008, pp. 1--10.

\bibitem{AODV-DSR}
C.~C. D.~S. Sunil Kr~Maakar, Manju~Khurana and D.~Srivastava, ``Performance
  evaluation of aodv and dsr routing protocols for flying ad hoc network using
  highway mobility model,'' \emph{Journal of Circuits, Systems and Computers},
  vol.~31, no.~01, 2022.

\bibitem{ZRP}
S.~B. Mohammed~Ahmed, S.~A. Hussain, L.~A. Latiff, N.~Ahmad, and S.~M. Sam,
  ``Performance evaluation of fanet routing protocols in disaster scenarios,''
  in \emph{2021 IEEE Symposium On Future Telecommunication Technologies
  (SOFTT)}, 2021, pp. 46--51.

\bibitem{JARMROUT}
C.~Pu, ``Jamming-resilient multipath routing protocol for flying ad hoc
  networks,'' \emph{IEEE Access}, vol.~6, pp. 68\,472--68\,486, 2018.

\bibitem{GLSR}
D.~Medina, F.~Hoffmann, F.~Rossetto, and C.-H. Rokitansky, ``A geographic
  routing strategy for north atlantic in-flight internet access via airborne
  mesh networking,'' \emph{IEEE/ACM Transactions on Networking}, vol.~20,
  no.~4, pp. 1231--1244, 2012.

\bibitem{RGR}
J.-D. M.~M. Biomo, T.~Kunz, and M.~St-Hilaire, ``Routing in unmanned aerial ad
  hoc networks: A recovery strategy for greedy geographic forwarding failure,''
  in \emph{2014 IEEE Wireless Communications and Networking Conference (WCNC)},
  2014, pp. 2236--2241.

\bibitem{QGeo}
W.-S. Jung, J.~Yim, and Y.-B. Ko, ``Qgeo: Q-learning-based geographic ad hoc
  routing protocol for unmanned robotic networks,'' \emph{IEEE Communications
  Letters}, vol.~21, no.~10, pp. 2258--2261, 2017.

\bibitem{RFLQGEO}
W.~Jin, R.~Gu, and Y.~Ji, ``Reward function learning for q-learning-based
  geographic routing protocol,'' \emph{IEEE Communications Letters}, vol.~23,
  no.~7, pp. 1236--1239, 2019.

\bibitem{QLCLRP}
C.~He, Q.~Wang, Y.~Xu, J.~Liu, and Y.~Xu, ``A q-learning based cross-layer
  transmission protocol for manets,'' in \emph{2019 IEEE International
  Conferences on Ubiquitous Computing and Communications (IUCC) and Data
  Science and Computational Intelligence (DSCI) and Smart Computing, Networking
  and Services (SmartCNS)}, 2019, pp. 580--585.

\bibitem{QLBR}
B.-S. Roh, M.-H. Han, J.-H. Ham, and K.-I. Kim, ``Q-lbr: Q-learning based load
  balancing routing for uav-assisted vanet,'' \emph{Sensors (Basel)}, vol.~20,
  no.~19, p. 5685, Oct 2020.

\bibitem{QLAR}
A.~Serhani, N.~Naja, and A.~Jamali, ``Qlar: A q-learning based adaptive routing
  for manets,'' \emph{2016 IEEE/ACS 13th International Conference of Computer
  Systems and Applications (AICCSA)}, pp. 1--7, 2016.

\bibitem{pathloss}
V.~V. Chetlur and H.~S. Dhillon, ``Downlink coverage analysis for a finite 3-d
  wireless network of unmanned aerial vehicles,'' \emph{IEEE Transactions on
  Communications}, vol.~65, no.~10, pp. 4543--4558, 2017.

\bibitem{nakagami}
I.~Atzeni, J.~Arnau, and M.~Kountouris, ``Downlink cellular network analysis
  with los/nlos propagation and elevated base stations,'' \emph{IEEE
  Transactions on Wireless Communications}, vol.~17, no.~1, pp. 142--156, 2018.

\bibitem{Ilimited}
W.~Liu, G.~Niu, Q.~Cao, M.-O. Pun, and J.~Chen, ``Particle swarm optimization
  for interference-limited unmanned aerial vehicle-assisted networks,''
  \emph{IEEE Access}, vol.~8, pp. 174\,342--174\,352, 2020.

\bibitem{my}
N.P.Sharvari, D.~Das, J.~Bapat, and D.~Das, ``Connectivity and collision
  constrained opportunistic routing for emergency communication using uav,''
  \emph{Computer Networks}, vol. 220, p. 109468, 2023.

\bibitem{energy}
W.~Heinzelman, A.~Chandrakasan, and H.~Balakrishnan, ``An application-specific
  protocol architecture for wireless microsensor networks,'' \emph{IEEE
  Transactions on Wireless Communications}, vol.~1, no.~4, pp. 660--670, 2002.

\bibitem{eng-eqs}
S.~J. Poudel~S, Moh~S, ``Residual energy-based clustering in uav-aided wireless
  sensor networks for surveillance and monitoring applications,'' \emph{Journal
  of Surveillance, Security and Safety}, vol.~2, no.~3, pp. 103--116, 2021.

\bibitem{engfly}
K.~Dorling, J.~Heinrichs, G.~G. Messier, and S.~Magierowski, ``Vehicle routing
  problems for drone delivery,'' \emph{IEEE Transactions on Systems, Man, and
  Cybernetics: Systems}, vol.~47, no.~1, pp. 70--85, 2017.

\end{thebibliography}

				\end{document}